# Title: Galactic Dynamics Using 1/r Force Without Dark Matter

**Author:** Martin Wen-Yu Lo[a*]

**Author Affiliation:**

[a]Jet Propulsion Laboratory, California Institute of Technology

***Corresponding Author***: Martin Wen-Yu Lo

Contact Information:

| | |
|---|---|
| Martin Wen-Yu Lo | 818-354-7169 (Voice) |
| JPL 168-200 | 626-429-9310 (Cell) |
| 4800 Oak Grove Dr. | 818-393-6962 (Fax) |
| Pasadena, CA 91109 | |

Martin.W.Lo@jpl.nasa.gov
http://www.gg.caltech.edu/~mwl

**Abstract:**

Dark matter, a conjectured substance not directly observable but which has tremendous mass, was proposed to explain why galaxies hold together and rotate faster at their edges than predicted by Newton's Inverse Square ($1/r^2$) Law of Gravity. Here we propose an alternative, an Inverse Law ($1/r$), which explains galactic morphology and rotation without dark matter. By varying initial conditions, the Inverse Law can systematically and easily generate realistic galactic formations including spirals, cartwheels (extremely difficult under Newtonian gravity), bars, rings, and spokes. This model can also produce filaments and void structures reminiscent of the large-scale structure of the universe. Newtonian gravity cannot do all this without dark matter. Occam's Razor suggests that at galactic scales, gravity should be $1/r$ and dark matter is unnecessary. This simple model with its self-organizing emergent properties, combined with dynamical systems theory, has broader implications. It may help us understand more complex systems.

**Main Text:**

Introduction

It is well known that Newton's Law of Gravity, when applied to galactic dynamics, leads to a large amount of missing matter that is not directly observable. Early in the 20$^{th}$ century, Oort (1) and Zwicky (2, 3) first discovered this phenomenon which Zwicky named, dark matter. Reuben & Ford (4) confirmed this. They observed that stars at the edge of galaxies move faster than predicted by Newton's $1/r^2$ Inverse Square Law of Gravity. This discrepancy is called the "Rotation Curve Problem" and currently is solved by adding dark matter around a galaxy. In this work, we propose an unorthodox but simple change of Newton's $1/r^2$ to $1/r$ for the weak-field approximation of gravity at the galactic scale. This approach solves several significant problems in galactic dynamics using the naïve (in the technical sense) direct simulation of the N-Body Problem in 2 dimensions. Surprisingly, it produced a series of realistic galaxy figures with flat rotation curves (which give correct rotation velocities at the edge of the galaxies) without dark matter. Both ring galaxies (Figure 1) and spiral galaxies (Figure 2) can be generated form this same model by simple changes to the initial conditions. Newtonian gravity cannot do this without dark matter as shown by Ostriker and Peebles (5). In particular, Toomre (6) showed the Cartwheel Galaxy can only be generated by the collision of two galaxies in Newtonian models (Compare our simulated Cartwheel Galaxy in Frame 100, Fig. 1, to the real Cartwheel Galaxy ESO 350-40 in the lower right corner of Fig. 1; both have an inner and outer ring connected by slightly curved spokes. Also compare our simulated spiral galaxy in Frame 30, Fig. 2, with the Barred Spiral Galaxy NGC1300 in the lower right corner of Fig. 2; both have two spiral arms, a bar, and a central core. Frame 110 of Fig. 2 shows the spiral galaxy evolved a typical ring surrounding the bar as well.

Use of the $1/r$ force for gravitation is not new. Fabris and Campos (7) and references therein pointed out numerous modifications to the theory of gravity recurrently in the literature. They cited no less than six different physical theories which added a $1/r$ term to gravity. These theories range from general relativity, string theory, quantum gravity, TeVeS theory, their own theory, and MOND theory (Modified Newtonian Dynamics, Milgrom (8, 9)). What is novel about our approach is the choice to retain only the $1/r$ term as the weak-field approximation of gravity to study dynamics at the galactic scale. This allowed us to simulate galaxies without dark matter or additional assumptions and constraints like relaxation techniques or needing the system to be in special equilibrium states. This greatly simplified the modeling, solves the Rotation Curve Problem, but, apparently, still captured some of the key global features of galactic dynamics. In the tradition of celestial mechanics, we call this model the "Restricted N-Body Problem" (RNBP).

RNBP galaxies undergo significant evolution in morphology with nearly periodic coherent structures. The periodic evolution of the morphology and the simplicity of the model suggest the use of dynamical systems theory to analyze the evolution of these coherent structures. In addition to giving insights into galactic dynamics, the RNBP may also provide a useful model to study more complex evolutionary processes with emerging structures such as in molecular dynamics or biological systems.

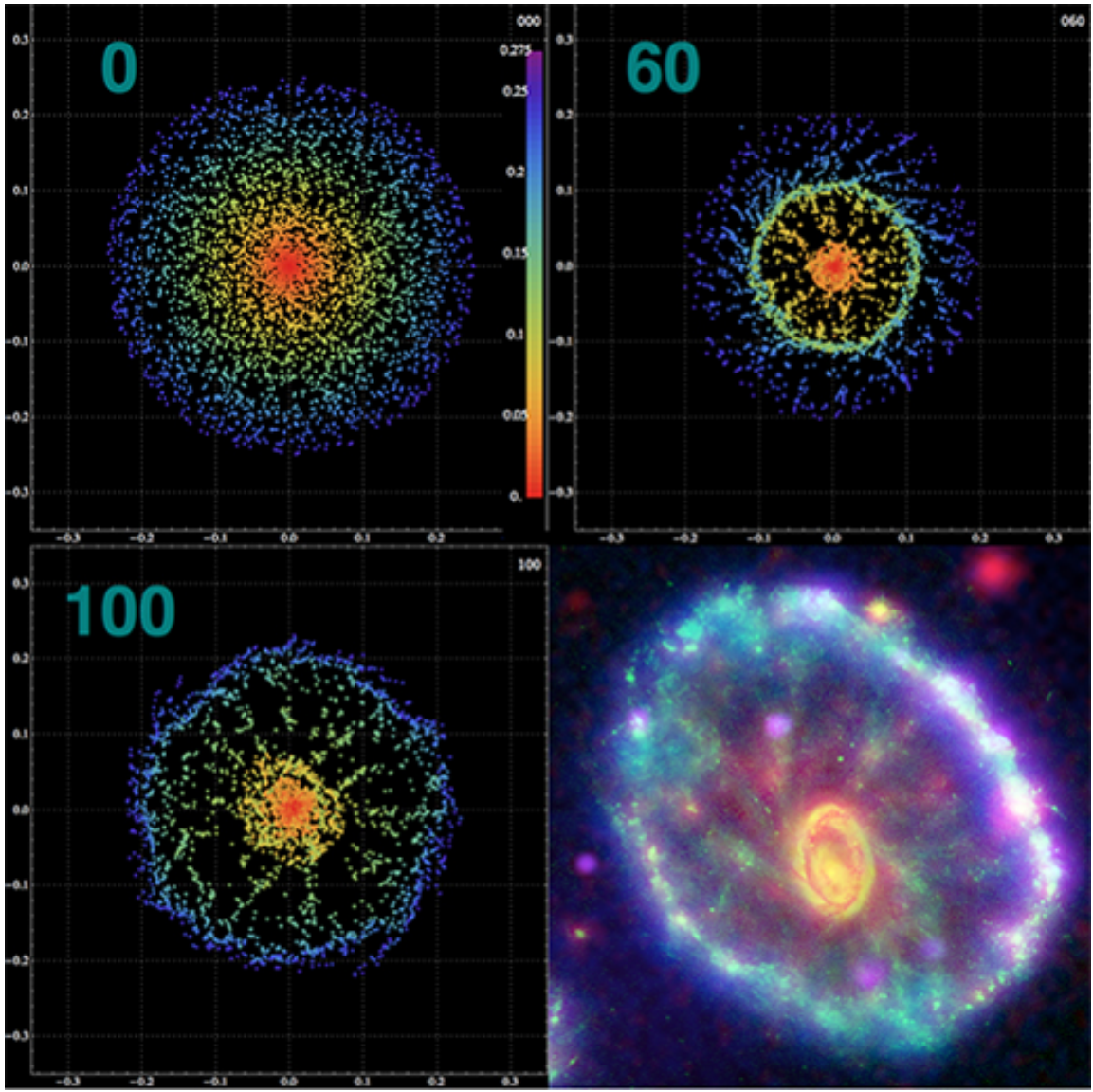

Figure 1. Case A (Table S1.) Simulated evolution of the Cartwheel Galaxy at Frames 0, 60, 100, compared with the Multispectral false color image of Cartwheel Galaxy ESO 350-40 (Credit: NASA/JPL-Caltech) in the lower right corner. Particles are color-coded by distance from galaxy center in Frame 0 by the color bar. Little mixing occurs as shown by the colors. Rings form by density waves travelling through the spokes which look like invariant manifolds of resonant periodic orbits in dynamical systems theory. See the additional figures in the Supporting Information attached, and the animation http://martinlo.com/Home/Cartwheel_Movie.html on-line.

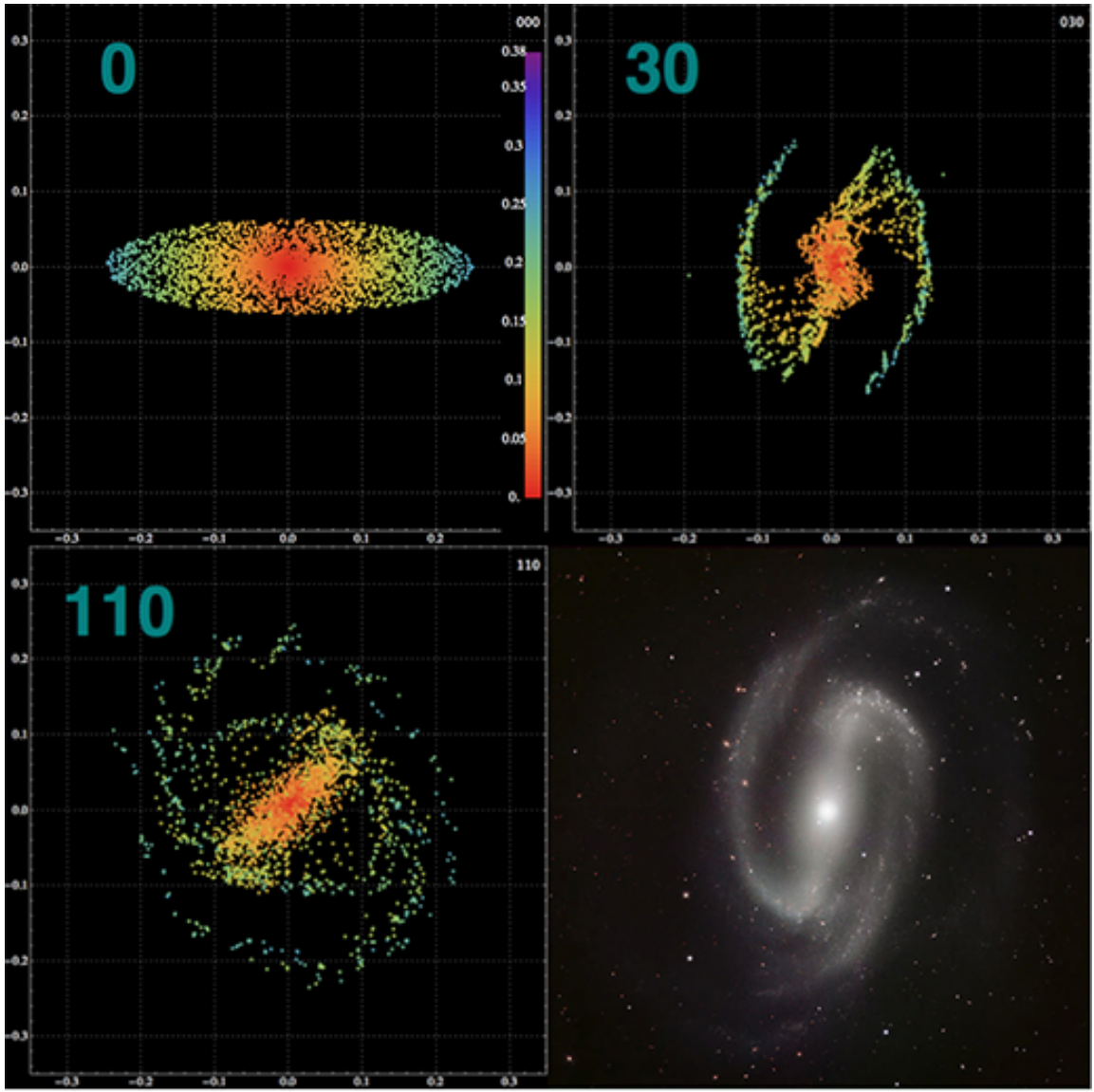

Figure 2. Case B (Table S1). Simulated evolution of the barred spiral galaxy with ring at Frames 0, 30, 110. Compare Frame 30 with the Barred Spiral Galaxy NGC1300 (Credit: ESO/P. Grosbøl). Particles are color coded by distance from galaxy center in Frame 0 by the color bar. Little mixing occurs as shown by the colors. A bar, ring, and two spiral arms develop with the bar spinning around the ring. See the additional figures in the Supporting Information attached, and the animation http://martinlo.com/Home/Spiral_Galaxy_Movie.html on-line.

**N-body Model of a Galaxy with 1/$r$ Force**

The equations of motion for the Restricted N-Body Problem (RNBP) are given by Eqs. (S1) and (S2) in the Supporting Information Appendix. Objects in the Supporting Information are number as S1, S2, etc. Our particles are assumed to be abstract point masses each of which should be thought of as a large cluster of stars and not as single stars. The 1/$r$ force law is a central force with potential equal to log($r$).

We restrict our attention to the 2-dimensional case to keep the model simple for analysis. Ring galaxies and spiral galaxies are typically very flat, so a 2D planar model is reasonable. Extension to 3 dimensions is straightforward by adding more components to the position, velocity, and force vectors.

To gain insight for interpreting the simulations, we examine the Two Body Problem first. This is a central force, so energy, momentum, and angular momentum are conserved. Hence, a Two Body orbit moves in a fixed orbital plane normal to the angular momentum vector (Pollard (10)). Bertrand's Theorem (Bertrand (11), Goldstein (12)) (implies that the 1/$r$ force has no periodic orbits besides circular orbits. Interestingly, all circular orbits in this potential have constant velocity,

$$v_c = \sqrt{\Gamma M} , \tag{1}$$

where $\Gamma$ is a new gravitational constant different from Newton's gravitational constant $G$ and $M$ is the mass of the central body. This is seen by equating centripetal acceleration with gravitational acceleration $v_c^2 / r = \Gamma M / r$ and simplify.

Another interesting property of the Two Body orbits of the 1/$r$ force is that they are all bounded in an annulus. They cannot collapse to 0 or escape to infinity! This is seen from the Two Body Energy, $E = v^2 / 2 + \Gamma M \log(r)$, which with algebra yields $r < \exp(E / \Gamma M) < \infty$. The 1/$r$ force is much stronger than the 1/$r^2$ force.

**Simulations Of Galaxies With 1/$R$ Force**

We use the symplectic Euler integrator with difference equations Eq. (S5) (Lambers and Lof (13), Lambers (14)). The units are non-dimensional and normalized. For simplicity, all particles have mass = 1. The initial conditions for the position and velocity of the $i^{th}$ particle are given in Eq. (2)m

$$\begin{aligned} \vec{r}_i^0 : x_i^0 &= \rho_i \cos(\theta_i), & y_i^0 &= \varepsilon \rho_i \sin(\theta_i), \\ \vec{v}_i^0 : \dot{x}_i^0 &= -V \rho_i \sin(\theta_i), & \dot{y}_i^0 &= V \rho_i \sin(\theta_i). \end{aligned} \tag{2}$$

The $\rho_i$'s are uniformly random numbers from 0 to 0.25; the $\theta_i$'s are uniformly random angles from 0 to $2\pi$. $V$ is the velocity scale factor around ~50. The shape factor $\varepsilon$ is set to 1 for a circular distribution (for ring galaxies) and $\varepsilon < 1$ for an elliptical distribution (for spiral galaxies). Since the uniform distributions are in polar coordinates, the effect in Cartesian coordinates is a distribution with particle density increasing towards the center of the galaxy at ( 0, 0 ). This simulates the greater mass at the center of galaxies. This turns out to be important for the evolution of the galaxies (see Case C below). Each particle is given a velocity transverse to its position vector from the center of the galaxy, ( 0, 0 ), to create a spin and avoid total collapse to the center. The particles are color coded by their initial distance from the galactic center so they can be tracked by their colors as the galaxy evolves. The gravitational constant $\Gamma = 0.05$ and the time step $\Delta t = 10^{-4}$. Recall that $\Gamma$ is a new gravitational constant different from Newton's $G$.

The shape of the initial conditions determines the morphology of the galaxy and its development. The shape factor $\varepsilon$ set to 1 gives a circular distribution for Case A which resulted in cartwheel galaxies. Setting $\varepsilon < 1$ gives an elliptical distribution for Case B which resulted in spiral galaxies. Hence, in the RNBP model, the morphology of galaxies are easily controlled by the density and shape of the initial distribution of particles.

The number of particles for the simulations ranged from $2500 \leq N \leq 4000$. The simulations were programmed in Matlab and F77. A simulation with 4000 particles and 2000 time steps requires ~10 minutes on a MacBookPro laptop. Due to the symplectic nature of the integrator, the energy variations (Eq. S3) of the simulations are always under 1%.

We examined four cases: (A) The Cartwheel Galaxy, (B) A Barred Spiral Galaxy, (C) A Uniformly Distributed Galaxy without a massive central core, and (D) Collision of Two Galaxies (A and B). Table S1 lists the parameters for each case. For the discussion of these cases to follow, please view the animations, additional figures and text in the Supporting Information for more details.

**Case A. The Cartwheel Galaxy:** The Cartwheel Galaxy (ESO 350-40) is a one of the rare ring galaxies that has a ring within a ring connected by slightly curved spokes. Zwicky (15) discovered it and called it "one of the most complicated structures awaiting its explanation on the basis of stellar dynamics". Under Newtonian gravity, Toomre (6) showed the only known method to simulate the Cartwheel Galaxy is by colliding two galaxies. However, with the $1/r$ force, the Cartwheel Galaxy evolves naturally from a circular distribution of randomly placed particles described by Eq. S5 and Table S1.

Fig. 1 provides 3 Frames from the Movie S1 to show the evolution of the Cartwheel Galaxy simulation. Frame 0 shows the initial circular distribution with a radius of 0.25 and a higher concentration of particles at the origin. The units are selected to keep the simulation within the square $[-1, 1] \times [-1, 1]$. The particles have a counterclockwise rotation. Initially, the entire galaxy contracts towards the center. In Frame 60, one can see the radius of the galaxy shrank from 0.25 to 0.2. The dark yellow and red center is much denser in Frame 60 where the greenish ring has now formed which grew out of the center from a density wave. In Frame 100 the ring has expanded to the edge of the galaxy and the color has now changed to blue. This shows that as the density wave pulses outwards, particles are not being carried along very far. In Frame 100, one can also see a new yellow ring formed closer to the center. In fact, a series of three rings can be seen at one time in the animation. Between the rings, spokes have formed, giving the Cartwheel Galaxy its distinct wheel shape. These spokes are highly reminiscent of invariant manifolds of unstable resonant orbits in the Restricted Three Body Problem. In fact, Athanassoula, Romero-Gomez, & Masdemont (16, 17) have shown, for a static galactic potential, that rings, bars, and spiral arms are indeed formed from invariant manifolds of periodic orbits. For our dynamic potential, these manifolds appear to be Lagrangian coherent structures.

**Case B. The Barred Spiral Galaxy:** Spiral galaxies are the quintessential galaxies we typically think of. For Case B, $\varepsilon = 0.25$, Frame 0 of Fig. 2 shows the initial conditions of the simulation. The galaxy first contracts from a radius of 0.25 to 0.2 as shown in Frame 30 of Fig. 2 where it has developed two spiral arms and a central bar. Traces of spoke-like structures can be seen connecting the arms to the central bar. From the animation, a density wave pulses from the center trying to form a ring. But due to the elliptical distribution, only two fragments of the ring are formed which became the arms. In Frame 110, an elliptical ring has formed to contain the bar. The bar actually spins inside this eye-shaped structure. The arms break up partially in Frame 110, but reform again in later Frames.

In both Cases A and B, the galaxy is constantly evolving and going through a series of nearly periodic structures. Recent work showed strong evidence that disk galaxies have under gone strong dynamical evolutions throughout the last 8 – 11 Gyr (Kassin et al. (18), Epinat , Amram, Balkowski, & Marcelin (19), Yang et al. (20), Neichel et al. (21)). The RNBP shows both ring and spiral galaxies undergo substantial evolutions which are nearly periodic. All of the intermediate shapes and structures produced by RNBP appear to be consistent with observed galaxies.

**Case C. Uniform Galaxy:** For Case C, the initial particles are uniformly distributed in Cartesian coordinates as shown in Figure 3, Frame 1, and Table S1. This example shows the significance of a massive central core to the morphology and evolution of the galaxy which we now describe.

An interesting observation in Frame 100 (Fig. 3), the particles clumped to form voids and filaments reminiscent of the large scale structure of the universe. Thereafter, this galaxy was never able to form a consistent massive core. At Frame 343, it produced a loosely organized galaxy with shadows of multiple spiral arms for a short duration. By Frame 2401, it evolved into an elliptical galaxy. More than 10,000 Frames later, it remained an elliptical galaxy, the shape of the oldest known galaxies. Case C suggests that a massive central core is critical for the formation of ring and spiral galaxies.

**Case D. Colliding Galaxies:** In Case D, we collided the Cartwheel Galaxy (4000 particles) with the Barred Spiral Galaxy (1500 particles). The results are shown in Figure 4 below. Frame 2 shows the two galaxies approaching one another. Frame 101 shows the two galaxies after multiple collisions but still separate. Frame 525 shows the two galaxies merged into an elliptical galaxy. The same software used to simulate the Cartwheel Galaxy and the Barred Spiral Galaxy was used to simulate the colliding galaxies. This shows how easy this kind of numerical experiments can be performed with the $1/r$ force because of its numerical stability.

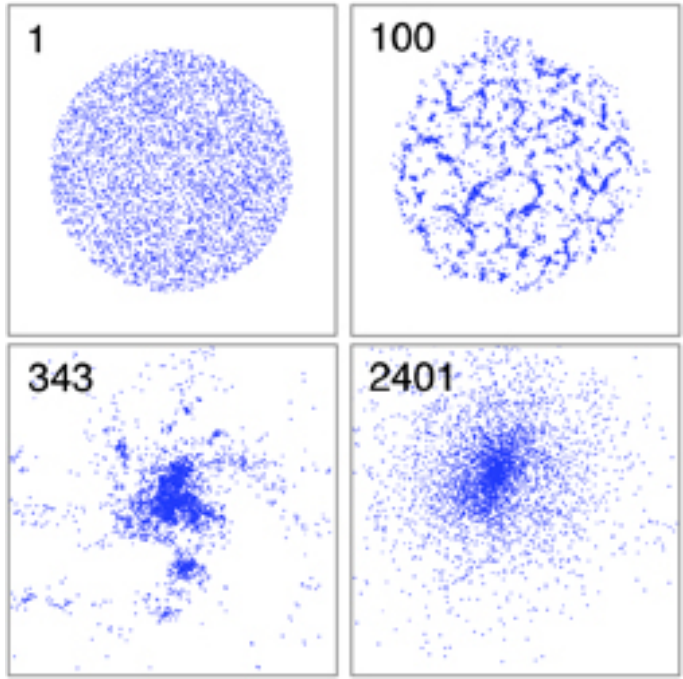

Figure 3. Case C (Table S1). Evolution of Uniform Distribution Initial Condition. Frame 1 shows the uniform random distribution initial condition. Here the rotation is slowed down: $V$=25. Frame 100 shows filament and void clumping reminiscent of the large scale structure of the universe. Frame 343 shows multiple spiral arms and a central core forming. Frame 2401 shows a final steady state elliptical galaxy. See http://martinlo.com/Home/Uniform_Galaxy_Movie.html on-line.

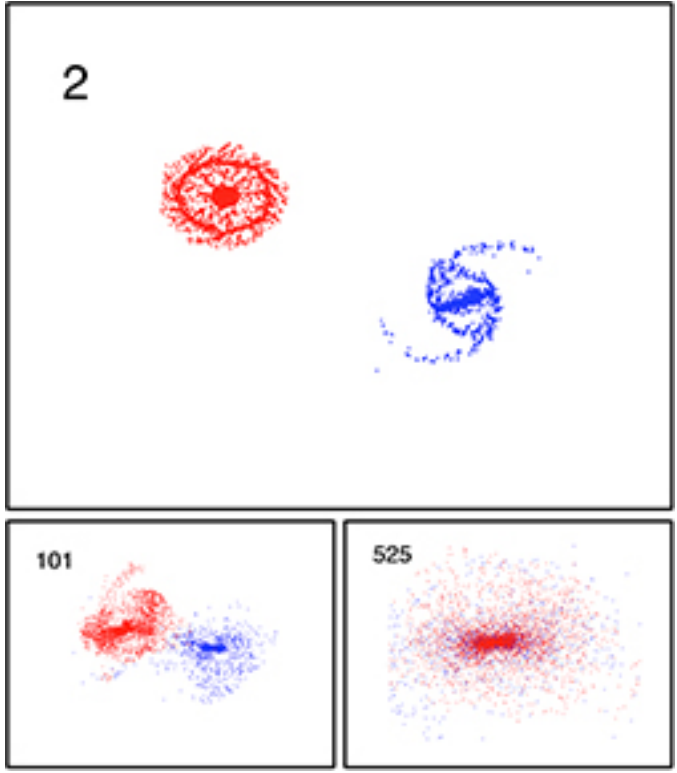

Figure 4. Case D (Table S1). Two Colliding Galaxies. Frame 2 shows the two galaxies approaching. Frame 101 is after several collisions. Frame 525 shows the two galaxies have combined and formed a steady state elliptical galaxy after multiple collision encounters. See the animation http://martinlo.com/Home/Galaxy_Collision_Movie.html on-line.

**GALAXY WITH 1/*r* GRAVITY FORCE HAS CONSTANT ROTATAION CURVE**

A key problem in galactic dynamics is the Rotation Curves of galaxies which describes the velocity of circular orbits around the galaxy center as a function of orbital radius. Keplerian circular orbits with radius *r* have velocity $v_k = \sqrt{GM/r}$, *M* is the mass of the central body. Thus, away from the center of the galaxy, circular velocity should become smaller. However, Rubin & Ford (4) observed the opposite: away from the center of the galaxy, they found the circular velocity to be nearly constant which produced a flat rotation curve. Recall the circular velocity for the Two Body Problem for the 1/*r* force is a constant, $v_c = \sqrt{\Gamma M}$. Surprisingly, for any N particles in the RNBP, the average circular velocity is still a constant, $v_c = \sqrt{\Gamma M}$, where now *M* = total mass of the galaxy.

The rotation curve for N bodies at a distance *r* from the center of the N-bides is given by

$$v_c(r) = \sqrt{r \frac{d\langle\Phi\rangle(r)}{dr}},$$

$$\langle\Phi\rangle(r) = \frac{1}{2\pi} \oint_{\|\vec{r}\|=r} \Phi(\vec{r}) \, r \, d\theta, \quad \Phi(\vec{r}) = \sum_{i=1}^{N} -\Gamma m_i \log\|\vec{r} - \vec{r}_i\| = \sum_{i=1}^{N} \Phi_i(\vec{r}), \quad (7)$$

where $\langle\Phi\rangle(r)$ is the azimuthal average of the log($r$) gravitational potential of the galaxy at distance $r$ from the center of the galaxy (Tremaine (22), Binney & Tremaine (23)). We showed in the Supporting Information Appendix that the rotation curve can be reduced to two definite integrals which by symmetry and cancellations reduces to the constant rotation curve, $v_c(r) = \sqrt{\Gamma M}$.

The significance of our calculations is that they are valid for *ANY DISTRIBUTION* of $N$ particles in *ANY CONFIGURATION* during any time step of the simulation under the 1/$r$ force. Thus the rotation curve for the 1/$r$ force and the RNBP model is always a constant $\sqrt{\Gamma M}$ and dark matter is not needed. This solves the Rotation Curve Problem.

Of course, the actual rotation cruves of galaxies are not flat towards the center of the galaxy. Since near the center of the galaxy, the assumption of the far-field approximation of gravitation fails, so one should not expect the 1/$r$ force to be working in this regime. The transition from 1/$r$ to 1/$r^2$ force is an unsolved problem in fundamental physics beyond the scope of this paper.

As a corollary, since we are able to measure the rotation curves of real galaxies, this suggests a potentially new approach to estimate the total mass $M$ of a galaxy. Of course, since real rotation curves are not constant near the center of galaxies where most of the mass of the galaxy is concentrated, more careful thought is needed for estimating the total mass $M$ of galaxies this way.

**THE log($r$) POTENTIAL SATISFIES GAUSS' LAW IN 2D BUT NOT IN 3D**

The log($r$) potential is harmonic in 2D but not in 3D. This means that it satisfies Gauss' Law in 2D, but not in 3D. This has some significant consequences. Gauss' Law states that the Newtonian gravitational flux, $F_{Newton}(S)$, through any closed surface, $S$, is proportional to the total mass, $M$, enclosed by $S$. Assume $S$ is a sphere of radius $r$ enclosing the point mass $M$ at the center of $S$. Eq. (8) gives the flux for the general force laws, 1/$r^n$, n=1, 2, 3, …

$$F(S) = \int_S \frac{\Gamma M}{r^n} r^2 d\Omega = 4\pi r^{2-n} \Gamma M \qquad (8)$$

where $\Omega$ is the solid angle of the unit sphere and $r^2 d\Omega = dS$ is the infinitesimal area element of the sphere $S$ of radius r. For Newtonian gravity, n=2 and $F_{Newton}(S) = 4\pi\, GM$ is a constant. But, for the log($r$) potential, n=1 and $F_{\log(r)}(S) = 4\pi rGM$ depends on $r$. In fact $F_{\log(r)}(S)$ depends on the size and shape of the surface $S$, and $F_{\log(r)}(S) \neq 0$ even when $S$ does not contain the mass $M$! This suggests potentially a way to estimate the distribution of matter within a galaxy by measuring the flux $F_{\log(r)}(S)$ and using inverse methods. Thus, here on Earth at the far edge of the Milkyway, we are able measure a non-zero $F_{\log(r)}(S)$ which may yield some information about the mass distribution within the Milkyway. Or by measuring $F_{\log(r)}(S)$ in a small volume in the direction of another galaxy like M31, we may be able to extract information about its mass and distribution.

**CONCLUSIONS AND FUTURE WORK**

In this paper we proposed the Restricted N-Body Problem (RNBP) with $1/r$ force for the far field approximation of gravity at the galactic scale. This enabled a naive direct N-Body simulation of galaxies in 2D with a flat rotation curve without dark matter. Simulations of 2D RNBP galaxies produced realistic looking galaxies that do not require assumptions of equilibrium states or relaxation techniques. The RNBP model is able to produce both the Cartwheel Galaxy and Barred Spiral Galaxies by simple variations in the initial conditions. We also showed the RNBP does not satisfy Gauss' Law in 3D. This suggests potentially new methods to measure the total mass and distribution of matter in galaxies.

The RNBP has many directions for future work. The most important is to verify and understand the fundamental physics of $1/r$ gravity and how gravity scales from $1/r^2$ at the scale of the solar system to $1/r$ at the galactic scale and beyond. The simplicity of RNBP allows for a dynamical systems analysis of the evolution and structures of galaxies. For example, the density waves creating the rings, spokes, spiral arms and bars appear, from the simulations, to be coherent structures controlled by invariant manifolds and symmetry properties of the initial conditions. This approach may provide a systematic theory for the formation, evolution, and classification of galactic structures. The RNBP from N=2, 3, 4 to N >> 1 provide a new family of classical mechanics problems which have somehow escaped the attention of researchers (see Saari (24, 25)). Every theorem in Newtonian celestial mechanics could be reinterpreted under the $1/r$ force.

Finally, the RNBP is perhaps the simplest of N-Body models with evolutionary and emerging properties. Moreover, it is numerically robust and highly stable. Simple mathematical models like the Three Body Problem or Henon's Map have played major roles in the development of science. Given the success of dynamical systems theory in the analysis of galaxies with static potentials by Athanassoula, Romero-Gomez, & Masdemont (17), similar approaches to study the RNBP will help us unravel the complex dynamics and controls behind the morphogenesis of the RNBP. Seeing the galactic structures emerge and evolve in the RNBP animations, one can imagine chemcial or biological systems undergoing similar dynamics and evolutions. The insights gained from studies of the much simpler RNBP will provide a foundation and stepping stones for approaching other N-Body systems with more subtle forces and more complex dynamics. In addition to contributions to the scientific problems, the RNBP may also contribute to computational algorithms, simulations, and visualizations involving N-Bodies because of its simplicity, robustness and stability.

**Acknowledgments.** This research was conducted at the Jet Propulsion Laboratory, California Institute of Technology under a contract with the National Aeronautics and Space Administration. Funding was provided by the NASA ROSES Applied Information Systems Research (AISR) Program 2009-2012. The JPL Strategic University Research Program provided a summer student grant in 2010. Special thanks to Joseph Bredekamp, Program Manager of AISR, for his long-term support. Thanks to Pedro Llanos for pointing out Lambers & Lof (13) to me. Thanks to William Wu, JPL, for the Mathematica animation software, and to William Wentzel for editing the animations and titles. Thanks to following colleagues for their help: Scott Tremaine (22, 23) for explaining how to compute the rotation curve for an N-Body simulation, Donald Saari (24, 25) for providing references to some of his work on the $1/r$ force law. Thanks to Alan Barr, Elizabeth Kay-Im, Andrew Lo, Cecilia Lo, Jay Cee Pigg, Paul von Allmen, William Wu for their reviews of the manuscript.

**Author Contributions.** Martin Lo: simulations, analysis, software, writing. This research originated from a serendipitous reference to a homework problem on the Direct Newtonian N-Body Problem (Lambers & Lof (13)) that Lo's student Pedro Llanos came across, which was the basis for earlier simulation experiments (Lo & Llanos (26), Llanos (27)). Later, after further analysis, Lo discovered that Lambers & Lof (13) made an error (Lambers (14)) in which they inadvertently proposed a force law of $1/r$ instead of $1/r^2$. Lo's attempt to resolve the apparent contradiction between this error and the realistic simulations of the $1/r$ force resulted in this paper.

**References**

1. Oort, J. (1932) The force exerted by the stellar system in the direction perpendicular to the galactic plane and some related problems, *Bull. Astron. Inst. Netherlands*, VI, No. 238, pp. 249-287.
2. Zwicky, F. (1933) Die Rotverschiebung von extragalaktischen Nebeln, *Helvetica Physica Acta* 6: 110–127. See translation in Zwicky, F. (2009) *Gen. Relativ. Gravit.*, 41:207-224.
3. Zwicky, F. (1937) On the masses of nebulae and of clusters of nebulae, *Astrophysical Journal* 86: 217-246.
4. Rubin, V. & Ford, W. K., Jr. (1970) Rotation of the Andromeda nebula from a spectroscopic survey of emission regions, *Astrophysical Journal* 159, pp. 379-403.
5. Ostriker, J., P. Peebles (1973) A Numerical study of the Stability of Flattened Galaxies: Or, Can Cold Galaxies Survive?, *Astrophyical Journal*, 186, pp. 467-480, December 1.
6. Toomre, A. (1978) Interacting systems, in *The Large Scale Structure of the Universe, (Proceedings of the Symposium, Tallin, Estonian SSR, September 12-16, 1977)*, Dordrecht, D. Reidel Publishing Co., pp. 109-116.
7. Fabris, J., Campos, J. (2009) Spiral galaxies rotation curves with a logarithmic corrected Newtonian gravitational potential, Gen. *Relativ. Gravit*. 41, pp. 93-104.
8. Milgrom, M. (1983) A modification of the Newtonian dynamics as a possible alternative to the hidden mass hypothesis, Astrophys. J. 270, pp. 365-370.

9. Milgrom, M. (1999) The modified dynamics as a vacuum effect, *Phys. Lett.* A 253, pp. 273-279.
10. Pollard, H., (1976) in *Celestial Mechanics*, MAA Carus Monographs No. 18, USA, pp. 1-10.
11. Bertrand, J. (1873) Théorème relatif au mouvement d'un point attiré vers un centre fixe, *C. R. Acad. Sci*. 77, pp. 849–853.
12. Goldstein, H. (1980) in *Classical Mechanics*, 2nd edition, Addison Wesley, Reading, MA.
13. Lambers, J., Lof, H. (2008) Assignment 3, Assignment N-body: The gravitational N-body Problem, CME212 Introduction to Large-Scale Computing in Engineering, Institute for Computational & Mathematcal Engineering, Stanford University, http://stanford.edu/class/cme212/.
14. Lambers, J. (2008) Questions Page (On error in Assignment 3), Feb 14, Feb 18, Feb 22, 2008, http://stanford.edu/class/cme212/. Accessed Nov. 2, 2011.
15. Zwicky, F. (1941) Contribution to Applied Mechanics and Related Subjects, Theodore von Kármen Anniversary Volume, Pasadena, CA, pp. 137.
16. Athanassoula, E., Romero-Gomez, M., Masdemont, J. (2009) Rings and Spirals in Barred Galaxies – I. Building Blocks, *MNRAS*, 394, pp. 67-81.
17. Athanassoula, E., Romero-Gomez, M., Masdemont, J. (2009) Rings and Spirals in Barred Galaxies – II. Rings and Spiral Morphology, *MNRAS*, 400, pp. 1706-1720.
18. Kassin, S., et al. (2012) The Epoch of Disk Settling: z~1 to Now, *Astrophys. J.,* pp. 758: pp. 106.
19. Epinat, B., P. Amram, C. Balkowski, and M. Marcelin (2010) *MNRAS* 401, pp. 2113-2147.
20. Yang, Y., et al. (2008) IMAGES. I, *A&A* 477, pp. 785-805.
21. Neichel, B. et al. (2008) IMAGES II., A&A 484, 1, 159-172.
22. Tremaine, S. (2012) private communications.
23. Binney, J. & Tremaine, S. (2008) Galactic Dynamics, Equation 2.165, Princeton University Press, Princeton.
24. Saari, Donald (1973). Improbability of Collisions in Newtonian Gravitational Systems II, *Transactions of the American Mathematical Society*, Vol. 181, July.
25. Saari, Donald (2010) MATHEMATICS AND ASTRONOMY: A JOINT LONG JOURNEY: Proceedings of the International Conference, *AIP Conference Proceedings*, Volume 1283, pp. 75-82.
26. Lo, M., Llanos, P. (2010) L5 Mission design & galaxy modeling, *JPL SURP Poster No. 10-SU-16, 2010 R&TD/DRDF/SURP Poster Conference*, JPL.
27. Llanos, P. (2011) Dynamics and morphology of galaxies, AIAA Region VI Student Conference, March 24-26, San Diego, CA.
28. Råde, L, Westergren, B. (2004) Mathematics Handbook for Science and Engineering, Springer Verlag, , Berlin, Integral No. 255 & 256, pp. 171.
29. Pauli, W. (2000) Pauli Lectures on Physics, Volume I, Electrodynamics, Dover Press, Mineola, NY.

# Supporting Information For Manuscript:

## Title: Galactic Dynamics Using 1/r Force Without Dark Matter

Author: Martin Wen-Yu Lo

Jet Propulsion Laboratory, California Institute of Technology

**Contact Information:**

    Martin Wen-Yu Lo                      818-354-7169 (Voice)
    JPL 168-200                           626-429-9310 (Cell)
    4800 Oak Grove Dr.                 818-393-6962 (Fax)
    Pasadena, CA 91109
    Martin.W.Lo@jpl.nasa.gov
    http://www.gg.caltech.edu/~mwl

I. Supporting Information Appendix

1. Supporting Information Appendix Introduction
   Figure S1 and caption summarizes entire paper.

2. Supporting Methods
   N-Body Model Of A Galaxy With $1/r$ Force
   Numerical Model Of Galaxies With $1/r$ Force

3. Supporting Table S1
   Simulation Cases

4. Supporting Figures S1-S10
   Figures For The Cartwheel Galaxy Simulation
   Figures For The Barred Spiral Galaxy Simulation

5. Supporting Discussion
   Galaxy With Log($r$) Potential Has Constant Rotation Curve
   The Log($r$) Potential Satisfies Gauss' Law In 2D But Not In 3D

6. Supporting References: All references refer to the references section in the main paper.

7. Lambers (13), Assignment 3: Assignment Nbody, from
   http://stanford.edu/class/cme212/, 2008.

8. Lambers (14), Questions Page (On error in Assignment 3), Feb 14, Feb 18, Feb 22, 2008, http://stanford.edu/class/cme212/.

   This webpage is no longer available since it is the course website and has been updated to the 2012 class at Stanford.

II. Supporting Movies S1-S4 Legends

    Movie S1: Cartwheel Galaxy & 1/$r$ Gravity Law
    Movie S2: Spiral Galaxy & 1/$r$ Gravity Law
    Movie S3: Uniform Galaxy & 1/$r$ Gravity Law
    Movie S4: Galaxy Collision & 1/$r$ Gravity Law

Movies located at: **http://martinlo.com/Home/RNBP.html**

# I. Supporting Information Appendix

## 1. Introduction

The Supporting Information Appendix contains the mathematical methods, equations of motion, numerical algorithms, additional figures, a table of the cases of galaxies simulated, and mathematical proofs of propositions in the main paper.

## 2. Supporting Methods

### N-Body Model Of A Galaxy With 1/r Force

The equations of motion for the Restricted N-Body Problem (RNBP) is given by Eqs. (S1) and (S2). Our particles are assumed to be abstract point masses each of which should be thought of as a large cluster of stars. The $1/r$ force law is a central force with potential equal to $-\log(1/r) = \log(r)$ [1]. Given two particles with mass $m_i$ and $m_j$ in positions, $\vec{r}_i$ and $\vec{r}_j$, the force of particle $m_j$ acting on $m_i$ with gravitational constant $\Gamma$ is given by $F_{ij}$:

$$F_{ij} = \frac{-\Gamma m_i m_j \vec{r}_{ij}}{r_{ij}^2},$$

$$\vec{r}_{ij} = \vec{r}_i - \vec{r}_j, \text{ where} \quad (S1)$$

$$\vec{r}_i = (x_i, y_i), \text{ a vector in the plane with real coordinates } x_i, y_i,$$

$$r_{ij}^2 = (x_i - x_j)^2 + (y_i - y_j)^2,$$

so $r_{ij} = \|\vec{r}_{ij}\|$ is the length of the vector, $\vec{r}_{ij}$. The force exerted on particle $m_i$ by the other N-1 particles is $F_i$:

$$F_i = \sum_{1 \le i < j \le N} F_{ij} = \sum_{1 \le i < j \le N} \frac{-\Gamma m_i m_j \vec{r}_{ij}}{r_{ij}^2} = -\nabla_i \Phi,$$

$$\Phi = \sum_{1 \le i < j \le N} \Gamma m_i m_j \log(r_{ij}), \quad (S2)$$

$$\nabla_i \Phi = \frac{\partial \Phi}{\partial x_i} + \frac{\partial \Phi}{\partial y_i}.$$

$\Phi$ is the potential of the N bodies. Note the gravitational constant $\Gamma$ here is different from Newton's gravitational constant, G. The total energy of the N-Body system is

$$E = \sum_{i=1}^{N} \frac{m_i v_i^2}{2} + \Phi. \quad (S3)$$

Extension to 3 dimensions is straightforward. These equations are valid in any dimension by adding more components to the vector $\vec{r}_{ij}$. For this paper, we restrict our attention to the 2-

---

[1] We use the expression $-\log(1/r)$ to show how it fits in with the general $1/r^n$ force laws which have potentials of the form $-1/r^{n-1}$ for $n > 1$. Ours is the special case n=1.

dimensional case to keep the model simple for ease of analysis and visualization. Ring galaxies and spiral galaxies are typically very flat so a planar model is reasonable. An interesting observation is that the log(*r*) potential is harmonic in 2D but not in 3D. The significance of this for galactic dynamics will be explained shortly.

To gain insight for interpreting the simulations, we first look at the Two Body Problem with 1/*r* force by setting $i = 1$, $j = 2$ in Equations (S1) and (S2). This is a central force, so energy, momentum, and angular momentum are conserved. Hence, a Two Body orbit moves in a fixed orbital plane normal to the angular momentum vector (see (Pollard (10)). Bertrand's Theorem (Bertrand (11), Goldstein (12)) implies that the log(*r*) potential has no periodic orbits besides circular orbits. Interestingly, all circular orbits in this potential have constant velocity,

$$v_c = \sqrt{\Gamma M},  \qquad (S4)$$

where $\Gamma$ is the new gravitational constant and *M* is the mass of the central body. This is seen by equating centripetal acceleration with gravitational acceleration $v_c^2 / r = \Gamma M / r$ and simplify.

Another interesting property of Two Body orbits of the log(*r*) potential is that they are all bounded in an annulus. There is no escape to infinity in this potential. This is seen from the Two Body Energy *E*, $v^2 / 2 = E - \Gamma M \log(r) \geq 0$, with algebra yields $r \leq \exp(E / \Gamma M) < \infty$. This is because the 1/*r* force is much stronger than the 1/*r*² force. For Two Body orbits with non-zero angular momentum, they are bounded below as well.

**Numerical Model Of Galaxies With 1/*r* Force**

Our numerical model for the RNBP uses the symplectic Euler integrator with difference equations at time step *n* given by (see Lambers & Lof (13), (Lambers (14)):

$$\vec{v}_i^{n+1} = \vec{v}_i^n + \Delta t \frac{F_i^n}{m_i},$$
$$\vec{r}_i^{n+1} = \vec{r}_i^n + \Delta t\ \vec{v}_i^{n+1}. \qquad (S5)$$

$\vec{r}_i^n, \vec{v}_i^n, F_i^n$, are the discretization of the position $\vec{r}_i$, velocity $\vec{v}_i = \dot{\vec{r}}_i$, and force $F_i$ of the $i^{th}$ particle at the $n^{th}$ time step, $n\Delta t$, n=0, 1, 2, ..., with $\Delta t$ the fixed time step. The problem is scaled so that the galaxy is evolving in a $[-1,1] \times [-1,1]$ square to simplify the visualization. The units are non-dimensional. For simplicity, all particles are assumed to have mass = 1. The initial conditions for the position and velocity of the $i^{th}$ particle are:

$$\begin{aligned}\vec{r}_i^0 : x_i^0 &= \rho_i \cos(\theta_i), & y_i^0 &= \varepsilon \rho_i \sin(\theta_i),\\ \vec{v}_i^0 : \dot{x}_i^0 &= -V\rho_i \sin(\theta_i), & \dot{y}_i^0 &= V\rho_i \sin(\theta_i),\end{aligned} \qquad (S6)$$

The $\rho_i$'s are uniformly random numbers from 0 to 0.25; the $\theta_i$'s are uniformly random angles from 0 to $2\pi$. *V* is the velocity scale factor around ~50. The shape factor $\varepsilon$ is set to 1 for a circular distribution (for ring galaxies) and $\varepsilon$ < 1 for an elliptical distribution (for spiral galaxies). Since the uniform distributions are in polar coordinates, the effect in Cartesian coordinates is a distribution with particle density increasing towards the center of the galaxy at $(0,0)$. This simulates the greater mass at the center of galaxies. Of course, it would be a simple modification to place a massive particle at the center to model the massive black hole at the center of galaxies. We leave this for future work. But we did look at Case C below with uniform density in Cartesian coordinates to see the effect of not having a massive core at the center of the galaxy. The initial velocity (Eq. S6) is perpendicular to the orbital radius to provide a rotation. Hence the system has a positive angular momentum which prevents total collapse during the course of our simulation. The gravitational constant $\Gamma = 0.05$ and the time step $\Delta t = 10^{-4}$. We remind the reader that $\Gamma$ is a fundamentally new gravitational constant different from Newton's G.

The number of particles for the simulations ranged from $2500 \leq N \leq 4000$. The simulations were programmed in Matlab and f77. A simulation with 4000 particles and 2000 time steps requires ~10 minutes on a MacBookPro laptop. Due to the symplectic nature of the integrator, the energy variations (Eq. S3) of the simulations are always under 1%.

## 3. Supporting Table

**Simulation Cases**

We examined four cases: (A) The Cartwheel Galaxy, (B) A Barred Spiral Galaxy, (C) A Uniformly Distributed Galaxy without a massive central core, and (D) Collision of Two Galaxies (A and B). To really appreciate these models, see the attached Supporting Information and the animations (http://martinlo.com/Home/RNBP.html) . Table S1 lists the parameters for each case.

| CASE NAME | $N$ | $\varepsilon$ | $V$ |
|---|---|---|---|
| A. Cartwheel Galaxy | 4000 | 1 | 50 |
| B. Barred Spiral Galaxy | 4000 | 0.25 | 50 |
| C. Uniform Galaxy | 4000 | 1 | 25 |
| D. Collision of 2 Galaxies: | | | 50 |
|     Barred Spiral Galaxy | 2500 | 0.25 | |
|     Cartwheel Galaxy | 4000 | 1.0 | |

Table S1. Parameters for simulations.

To track the particles and detect the amount of mixing as the galaxies evolve in Cases A and B, we color coded the particles by their initial distances from the center of the galaxy at time t=0 which is Frame 0 of each of the animations. A color bar for Case A, the Cartwheel Galaxy, is given in Figure S1, Frame 0. The color bar for Case B, the Spiral Galaxy, is given in Figure S2, Frame 0. At the start of each simulation, the red particles are at the center of the galaxy and the blue particles are at the outer edge of the galaxy. The figures of the simulations are time-tagged by the animation Frame Number at the top left corner of each frame. Depending on the simulation, the actual number of time steps $\Delta t$ between frames can vary from 1 to 20 for the simulations in Supporting Section.

## 4. Supporting Figures

**Figures For The Cartwheel Galaxy Simulation**

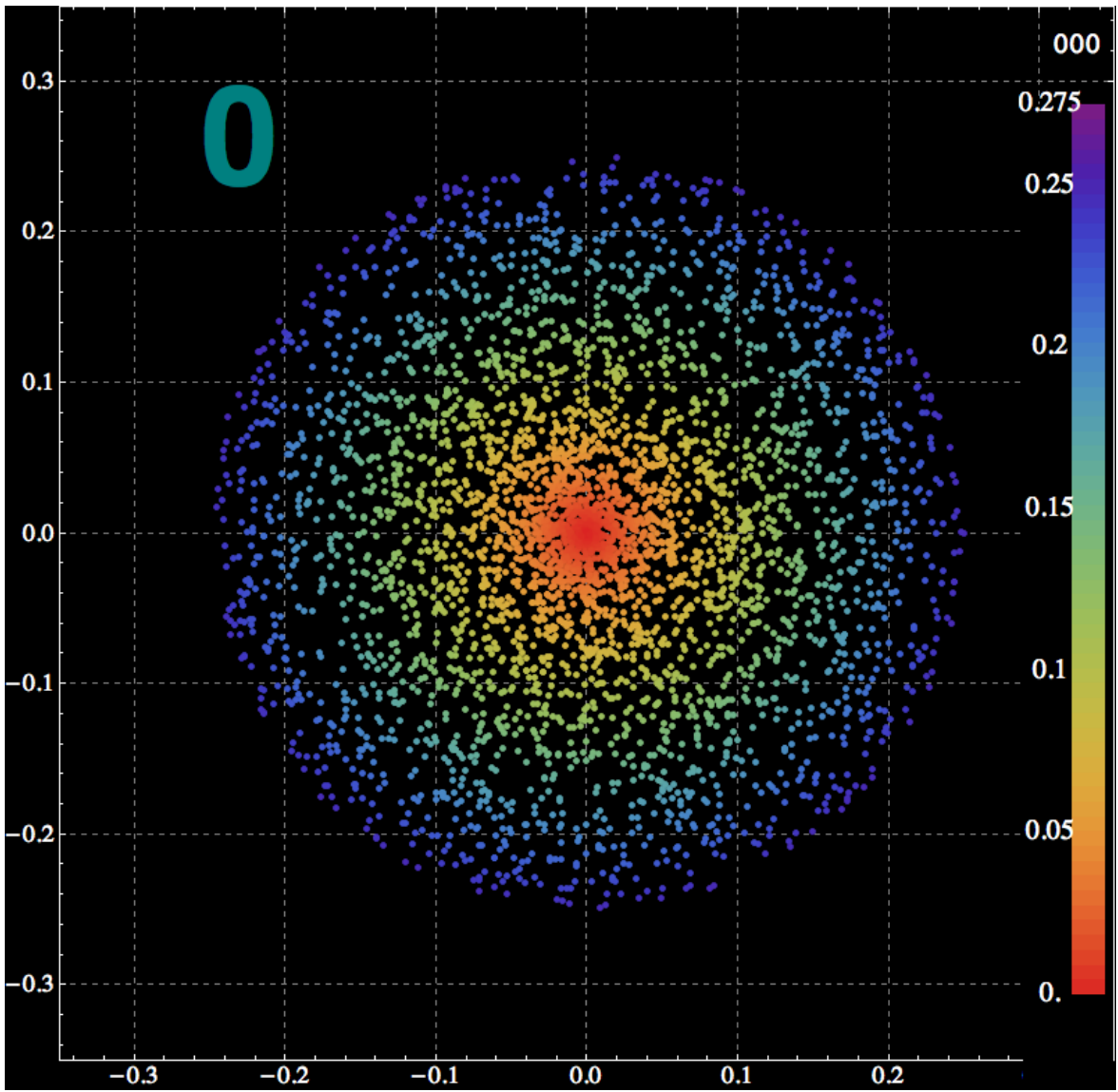

Figure S1. The initial random distribution of particles for the Cartwheel Galaxy simulation. The particle color is fixed according to the particle's initial distance form the center of the galaxy as shown in the color bar: red is at the center, blue is at the edge of the galaxy at time 0.

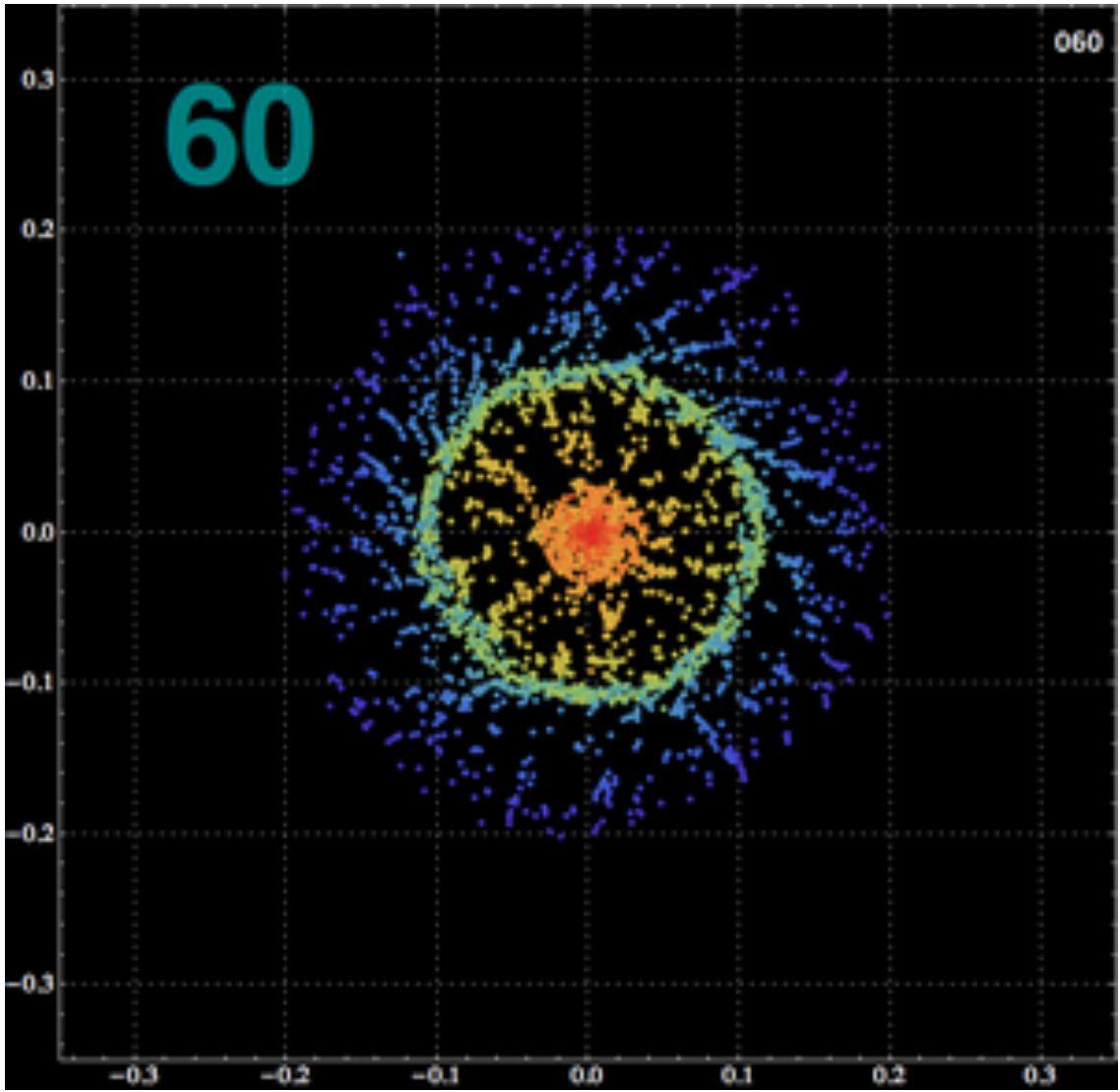

Figure S2. Frame 60 has a small yellow ring near the center and a green ring in the middle with spokes on either side of the green ring. Spokes connect the green ring to the yellow ring. The spokes look like the invariant manifolds of resonant orbits in the Circular Restricted 3 Body Problem. It is conjectured a similar dynamics is creating these coherent structures.

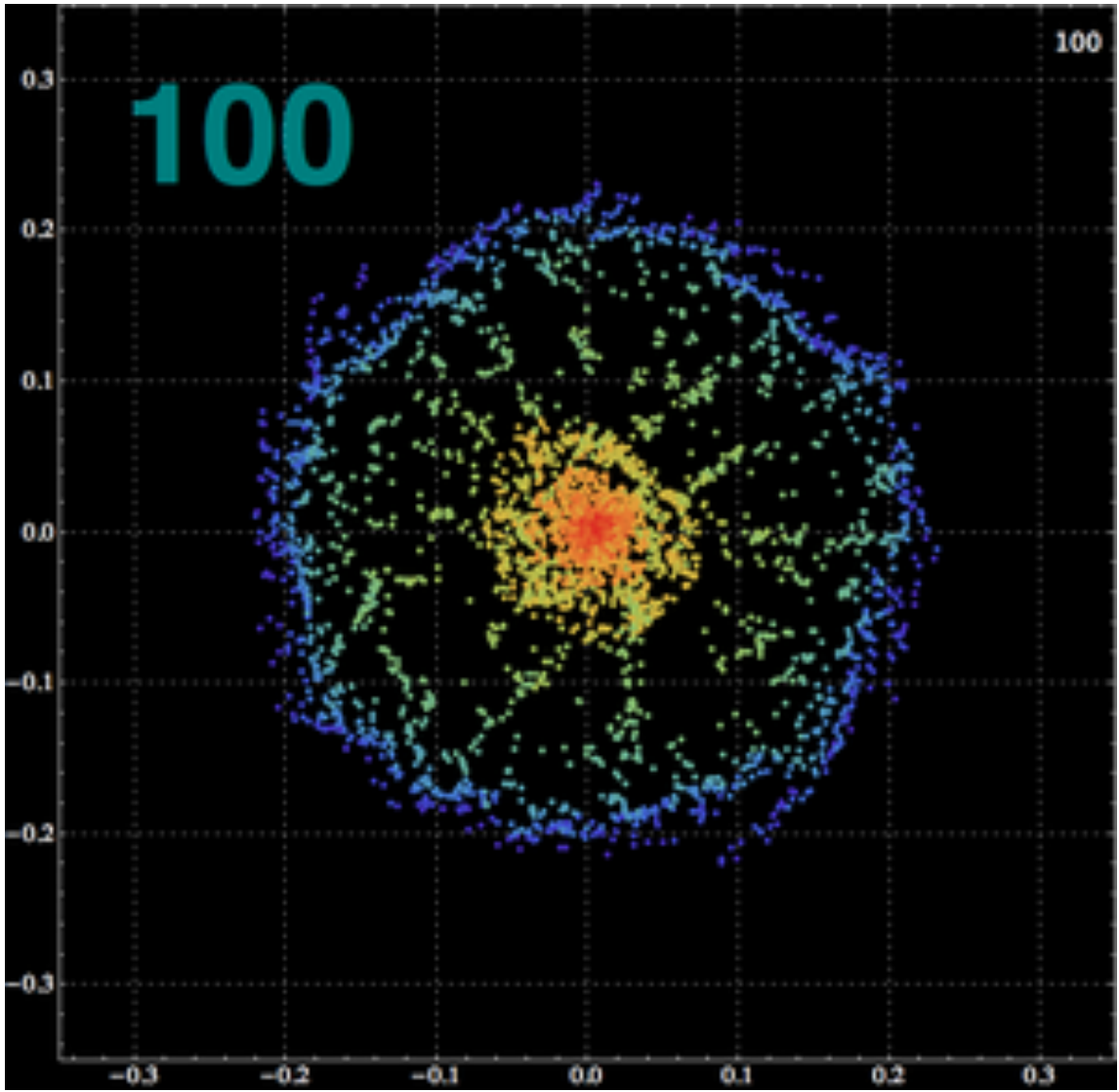

Figure S3. The ring of the density wave in Frame 60 which was green and located at half the radius of the galaxy has now expanded in Frame 100 to the edge of the galaxy and become blue particles. The green particles have now become the spokes connect the outer blue ring to the inner yellow ring. Whispy spiral arms outside of the blue ring look like invariant manifolds (whiskered tori) of resonant orbits in the CR3BP.

Figure S4. Animation of the Cartwheel Galaxy simulation using the log(*r*) potential in the

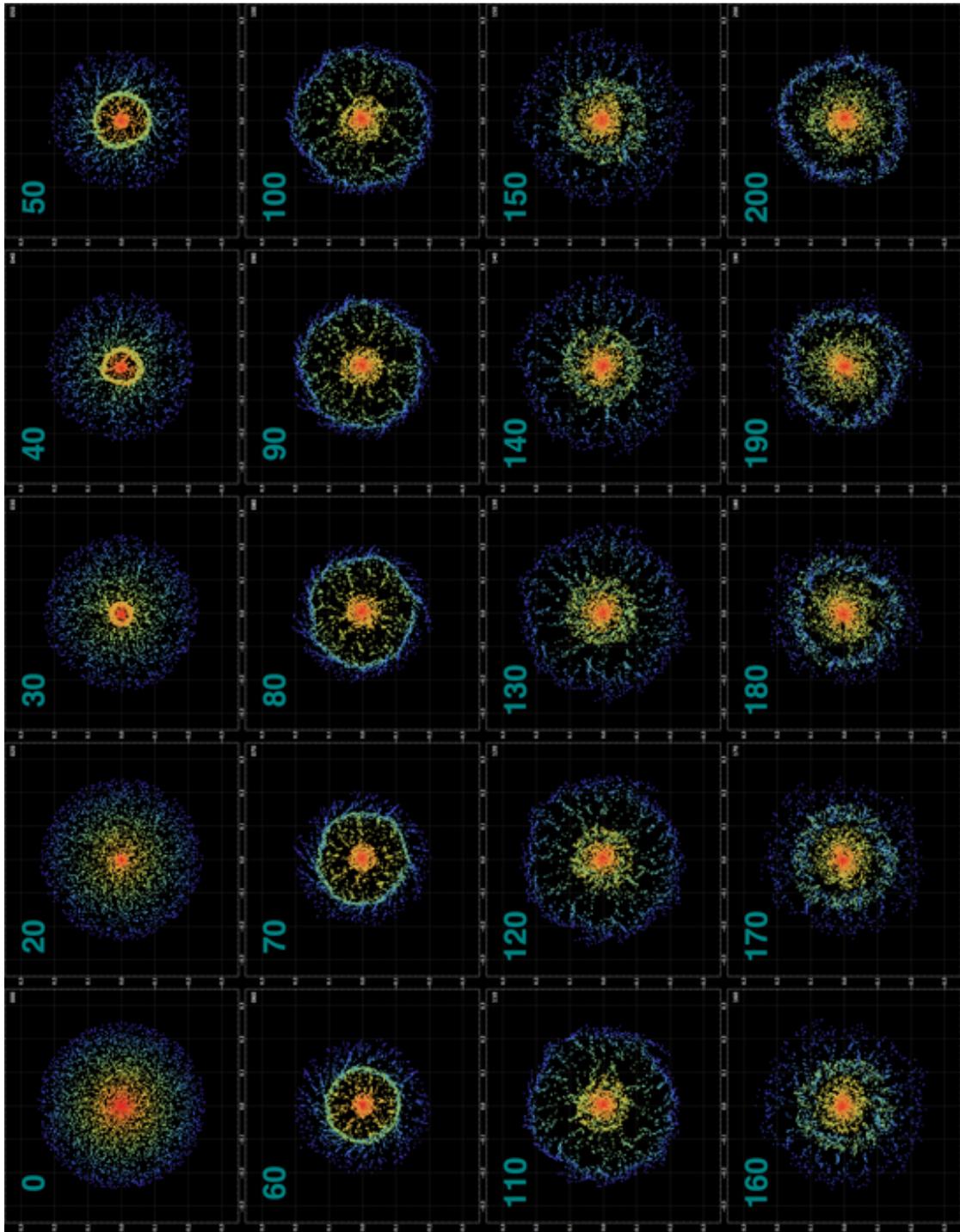

Supporting Information (see "The Cartwheel and Spiral Galaxy Movie"). See the main text for a description of the evolution of the galaxy. For greater details on Frame 60 and Frame 100 see Figs. S3 and S4.

**Figures For The Barred Spiral Galaxy Simulation**

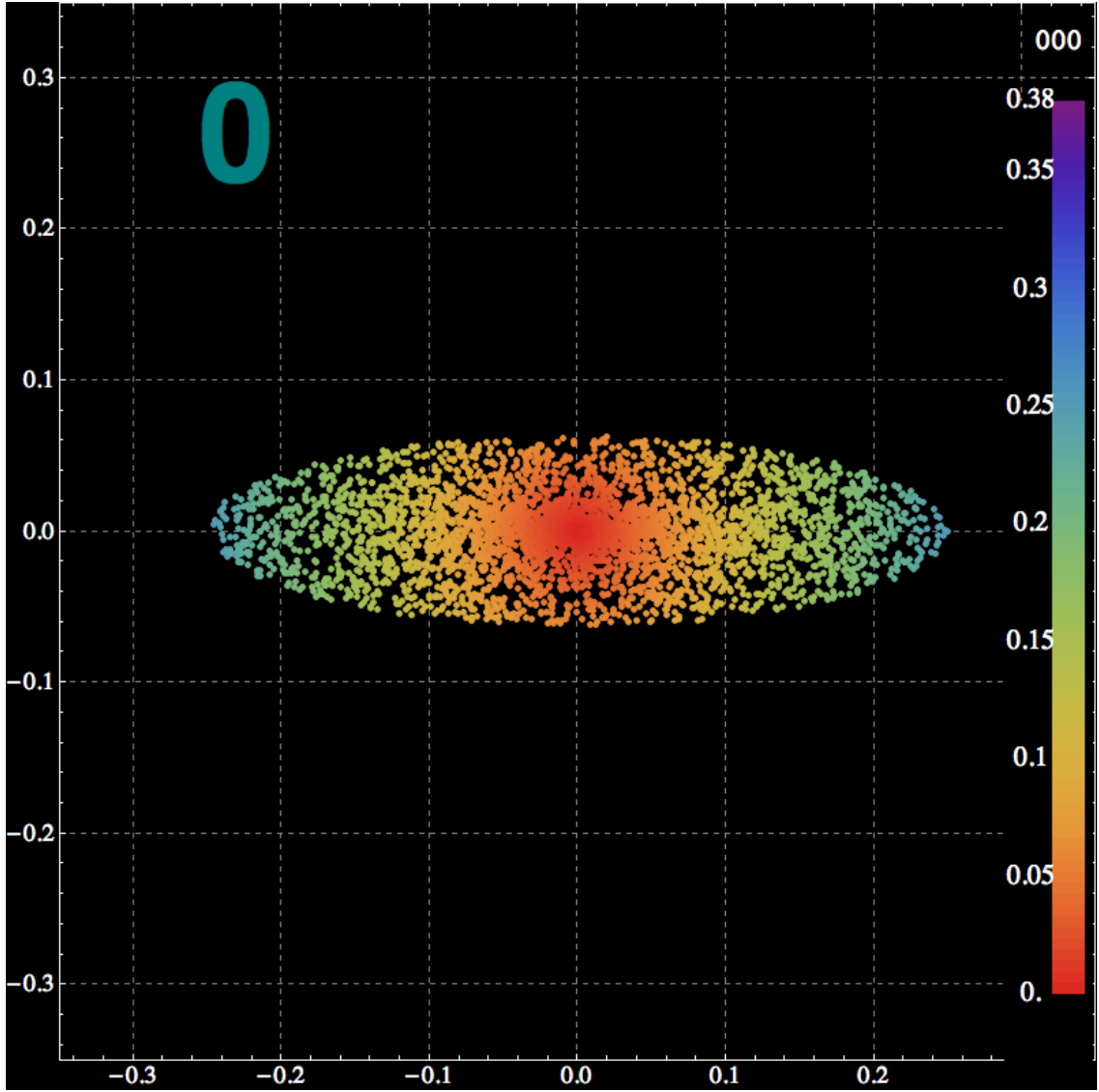

Figure S5. The initial random distribution of particles for the Barred Spiral Galaxy simulation with a shape factor ε = 0.25. The particle color is fixed according to the particle's initial distance form the center of the galaxy as shown in the color bar: red is at the center, blue is at the edge of the galaxy at time 0.

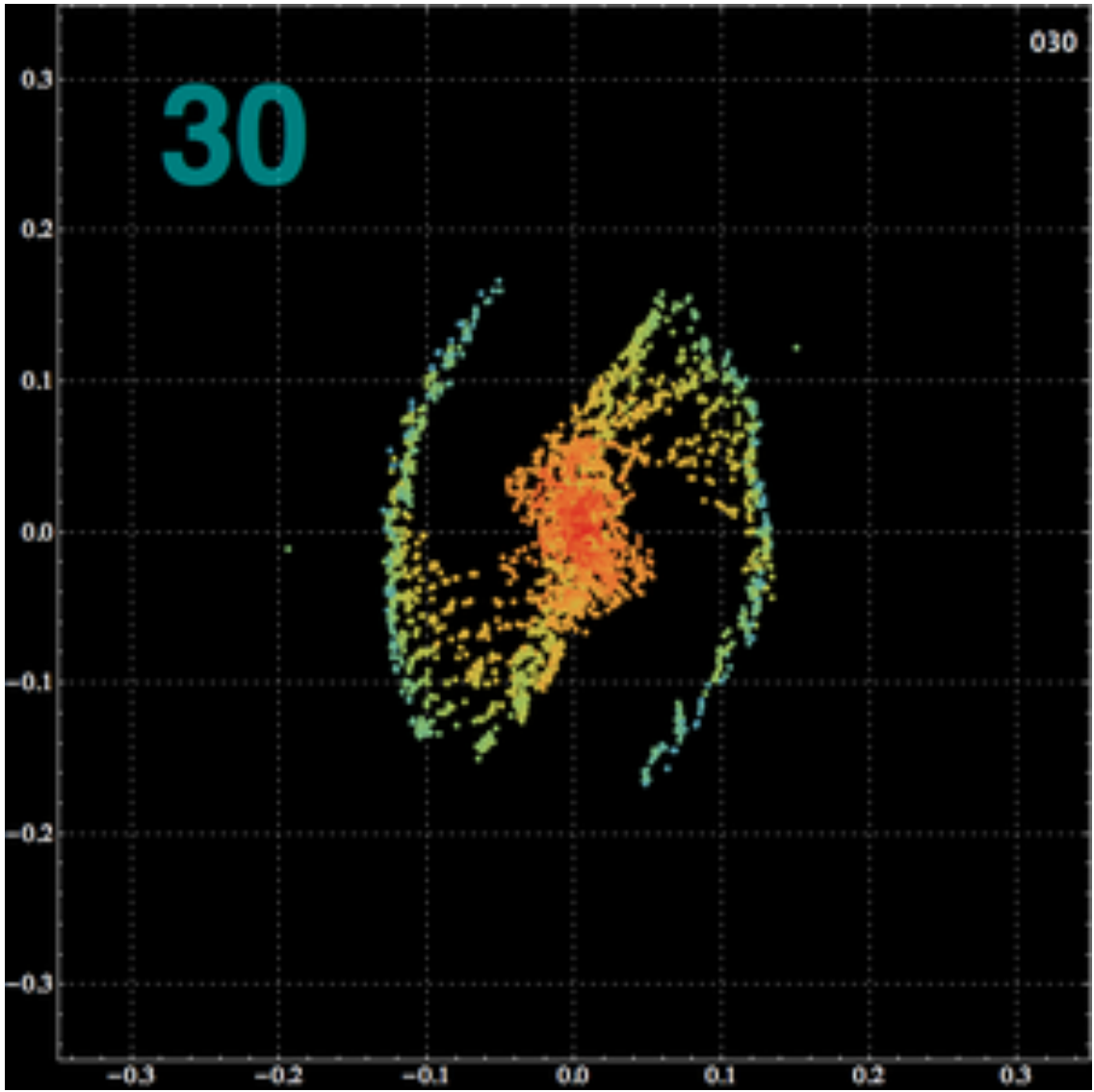

Figure S6. Frame 30 shows the early development of the Barred Spiral Galaxy simulation from the initial elliptical distribution in Frame 0, Fig. 8. This compares well with NGC1300 shown in Fig. 1.c and Fig. 1.d. Note the spoke like features connecting the outer edge of the spiral arms to the rectangular bar feature at the galactic core. This looks like a cartwheel galaxy with a partial ring and spokes.

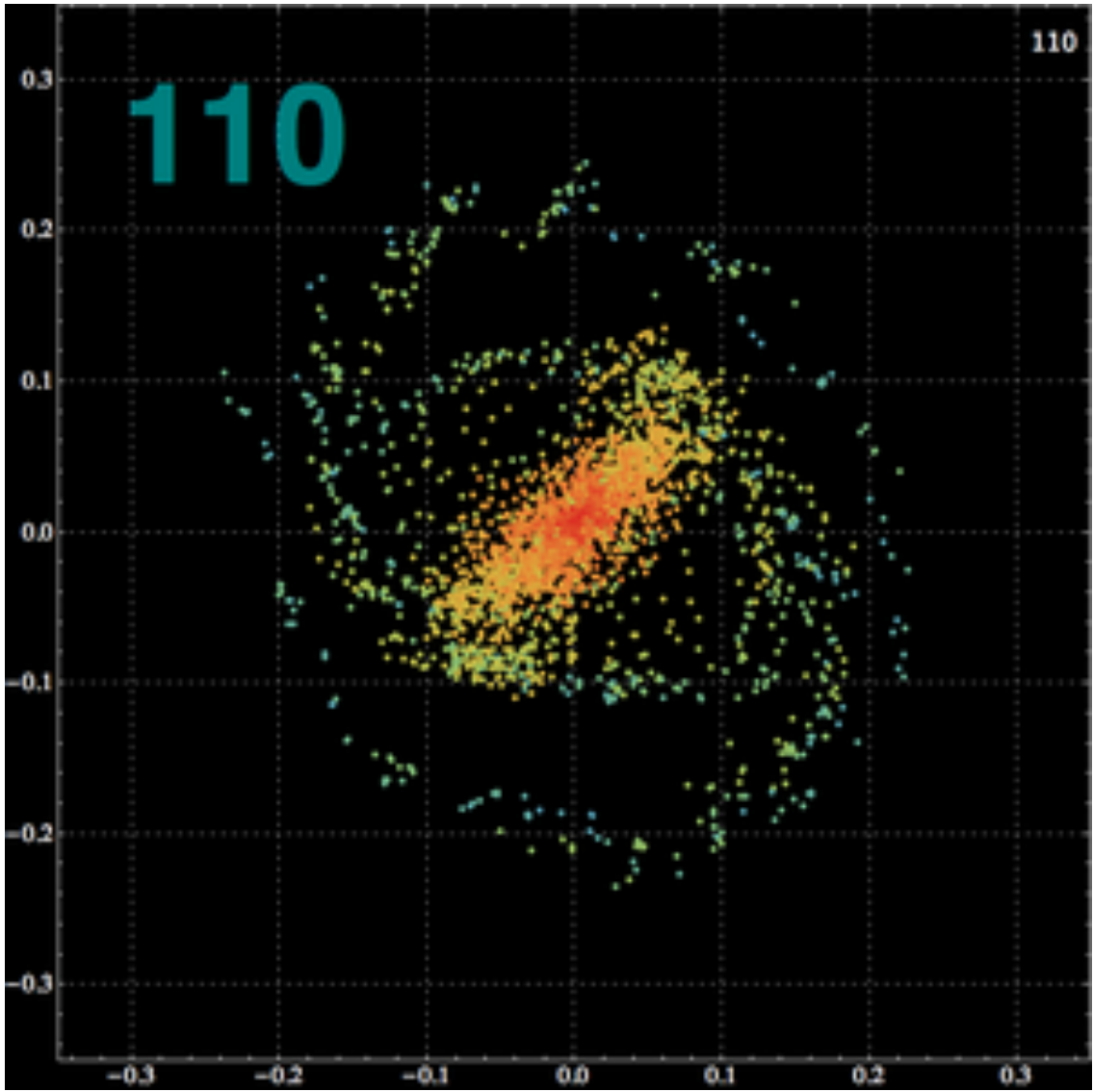

Figure S7. Frame 110 shows the spiral arms breaking up into spoke-like structures. These correspond to the whiskers on the outermost rings of the Cartwheel Galaxy. The bar in the center is changing phase with respect to the ring surrounding it as the galaxy evolves.

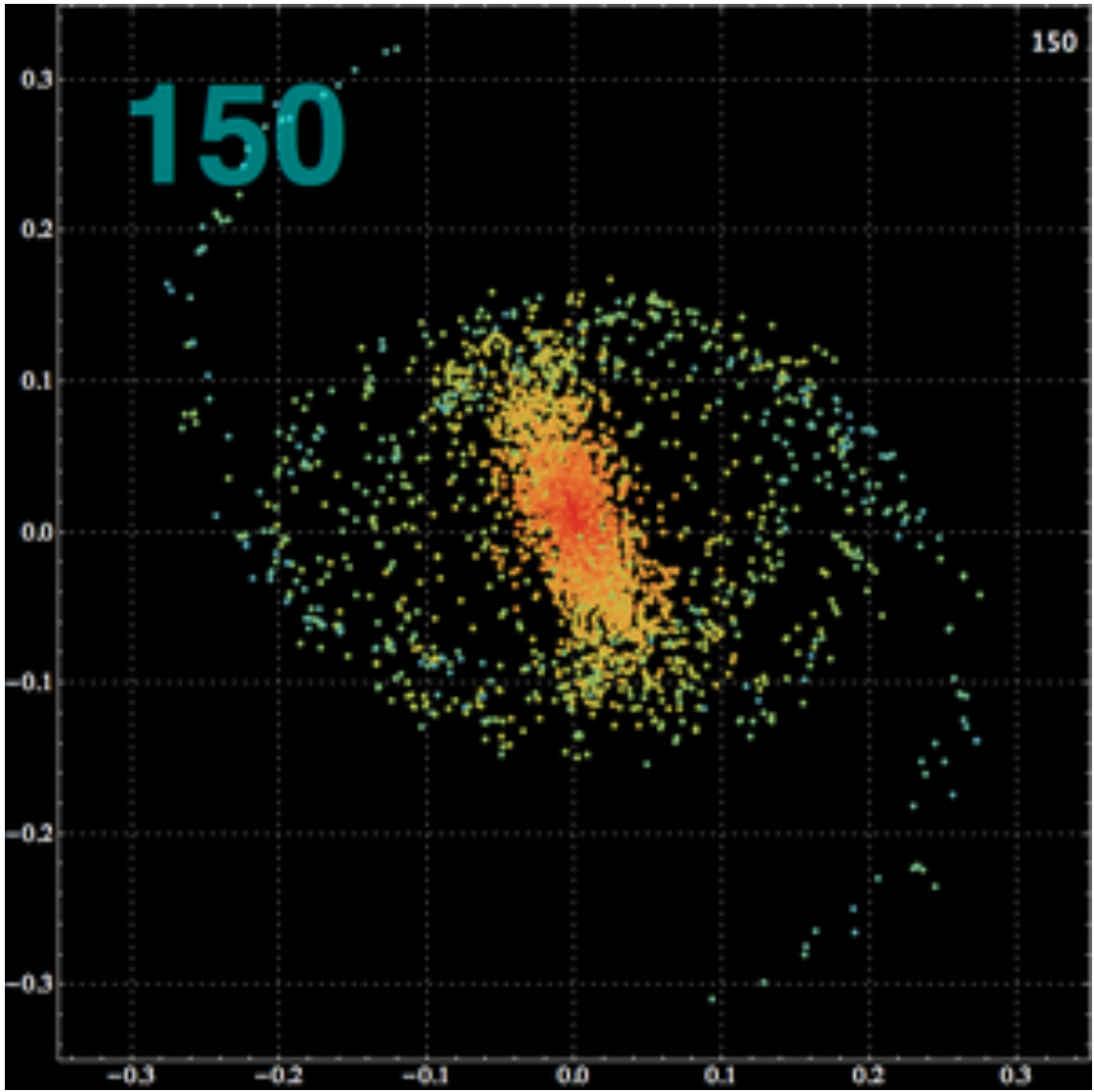

Figure S8. Frame 150 shows the two spiral arms have reconstituted from the whiskers in Frame 110. The bar is surrounded by a smaller circular ring within a larger elliptical ring. The phase of the bar within rings have moved counterclockwise from that of Frame 110.

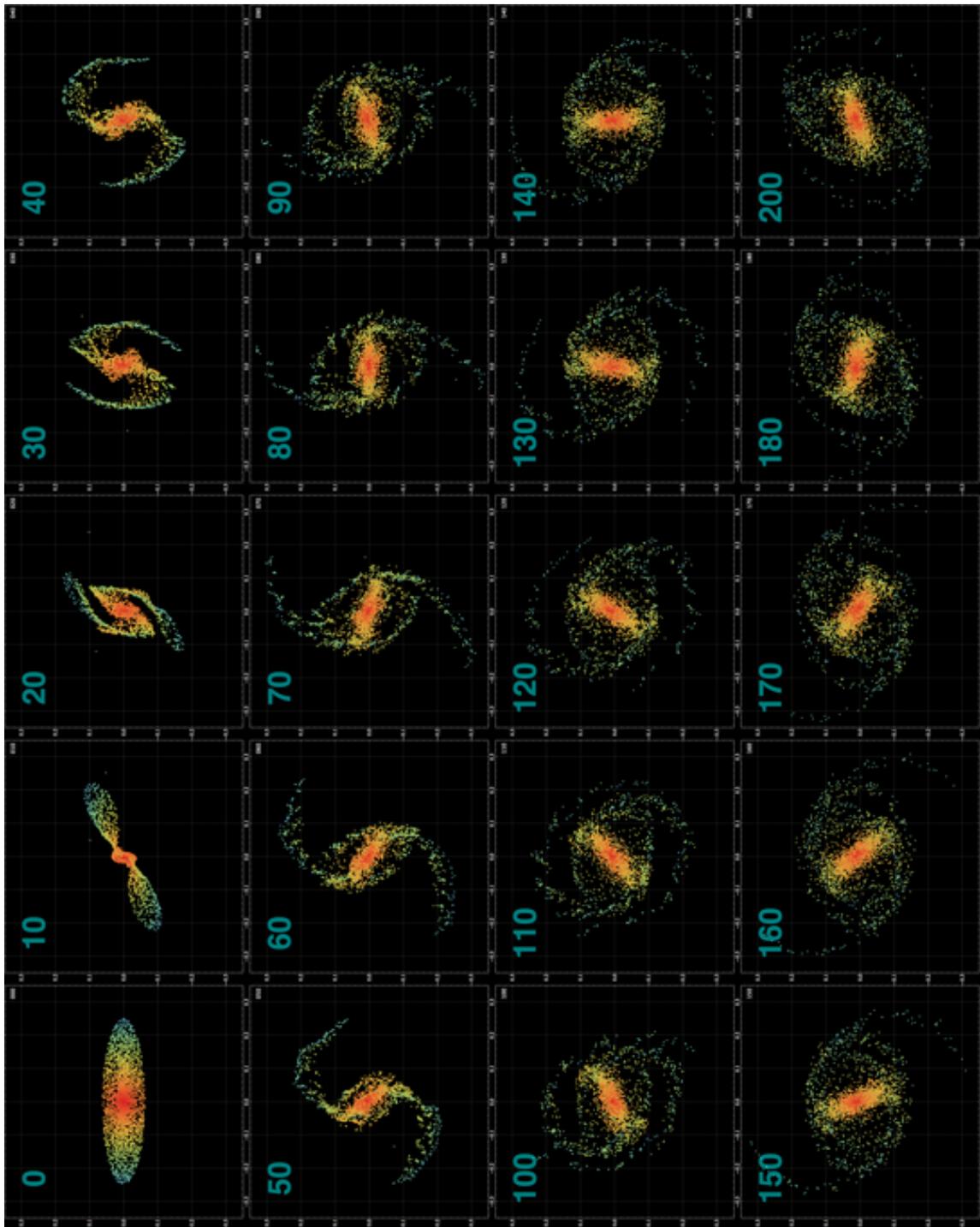

Figure S9. These 20 frames from the animation of the simulation shows how the Barred Spiral Galaxy evolves, starting from a randomly distributed set of particles in an elliptical shape with a rotation as initial conditions. Notice bars, spiral arms, rings emerge and evolves.

## 5. Supporting Discussion

**Galaxy With Log(*r*) Potential Has Constant Rotation Curve**

Having generated several simulated galaxies, a key question to address is the Rotation Curves of these galaxies. This is the curve describing the velocity of circular orbits around the center of the galaxy as a function of the orbital radius. Keplerian orbits have circular velocity of $v_k = \sqrt{\mu/r}$ as a function of its radius, $r$; thus the circular velocity should become smaller the further we are away from the center of the galaxy. However, this is not what Rubin & Ford (4) 1970 observed. Instead, as one goes away from the center of the galaxy, they found the circular velocity became a constant. This surprising result confirmed the previous observations of Oort (1) 1932 and Zwicky (2, 3) 1933 that there is missing matter in the galaxy, or Dark Matter as named by Zwicky. With the log(*r*) potential of the RNBP, we have already observed in Eq. (S2) that the circular velocity for the Two Body Problem is always a constant, $v_c = \sqrt{\mu} = \sqrt{\Gamma M}$. What about the rotation curve for each of our RNBP simulations? The surprising result we obtained is that even in the N-body case, the average circular velocity of the log(*r*) potential is still a constant, $v_c = \sqrt{\Gamma M}$, where now $M$ = total mass of the galaxy. This is totally unexpected since we are starting from a set of N particles distributed somewhat arbitrarily in space and then evolved in time. We provide a sketch of the calculations. Basically, the rotation curve is reduced to two definite integrals which by symmetry and cancellations reduces to the constant circular velocity $v_c = \sqrt{\Gamma M}$.

The rotation curve for N bodies at a distance $r$ from the center of the N-bides is given by

$$v_c(r) = \sqrt{r \frac{d\langle\Phi\rangle(r)}{dr}},$$

$$\langle\Phi\rangle(r) = \frac{1}{2\pi}\oint_{\|\vec{r}\|=r}\Phi(\vec{r})\,r\,d\theta, \quad \Phi(\vec{r}) = \sum_{i=1}^{N}-\Gamma m_i \log\|\vec{r}-\vec{r}_i\| = \sum_{i=1}^{N}\Phi_i(\vec{r}),$$

(S8)

where $\langle\Phi\rangle(r)$ is the azimuthal average of the log(*r*) gravitational potential of the galaxy at distance $r$ from the center of the galaxy (Tremaine (22), Binney & Tremaine (23) ).

To compute Eq. (S8), we assume the $\vec{r} \neq \vec{r}_i$ which is reasonable since there are only a finite number of particles which is a set of measure 0; then $\Phi(\vec{r})$ is well behaved at a fixed $r = \|\vec{r}\|$ so that we can exchange the integral and derivative to compute

$$\frac{d\langle\Phi\rangle(r)}{dr} = \frac{1}{2\pi}\oint_{\|\vec{r}\|=r}\frac{\partial\Phi(\vec{r})}{\partial r}\,r\,d\theta = \frac{1}{2\pi}\sum_{i=1}^{N}\oint_{\|\vec{r}\|=r}\frac{\partial\Phi_i(\vec{r})}{\partial r}\,r\,d\theta,$$

$$\Phi_i(\vec{r}) = -\Gamma m_i \log\|\vec{r}-\vec{r}_i\|.$$

(S9)

By linearity, we can compute each $i^{th}$ term under the summation in Eq. (S9) separately. Since we are averaging $\partial\Phi_i/\partial r$ over the circle of radius $r$, we can simplify the calculation by rotating the coordinates so that $\vec{r}_i = (r_i, 0)$ is on the X-axis. This enables us to break up the $i^{th}$ integral into two definite integrals of the form

$$r \int_0^{2\pi} \frac{\cos(a\theta)d\theta}{b + c\cos(a\theta)} - r_i \int_0^{2\pi} \frac{d\theta}{b + c\cos(a\theta)} \tag{S10}$$

These are integrable (see Råde (28)) and with some algebra yields

$$\frac{d\langle\Phi_i\rangle(r)}{dr} = \frac{\Gamma m_i}{r}. \tag{S11}$$

Substituting Eq. (11) into Eq. (9), we obtain the rotation curve of $N$ particles is a constant

$$v_c(r) = \sqrt{r \sum_{i=1}^{N} \frac{\Gamma m_i}{r}} = \sqrt{\Gamma M},$$

$$M = \sum_{i=1}^{N} m_i. \tag{S12}$$

This is the same as the constant circular velocity of the Two Body orbits of the $\log(r)$ potential.

The significance of this calculation is that Eqs. (9) – (12) is valid for *ANY DISTRIBUTION* of $N$ particles in *ANY CONFIGURATION* during any time step of the simulation under the $\log(r)$ potential. The rotation curve is always a constant $\sqrt{\Gamma M}$ and Dark Matter is not needed.

As a corollary, since we are able to measure the rotation curve of real galaxies which was how Dark Matter was rediscovered by Rubin & Ford (4), this offers potentially a new approach to estimate the total mass $M$ of a galaxy. Of course, since real rotation curves are not constant near the center of galaxies where most of the mass is concentrated, more careful thought is needed for estimating the total mass $M$ this way.

**The Log(*r*) Potential Satisfies Gauss' Law In 2D But Not In 3D**

The $\log(r)$ potential is harmonic in 2D but not in 3D. This means that while it satisfies Gauss' Law in 2D, it does not satisfy Gauss' Law in 3D. We will give a heuristic explanation of this observation and its consequences. A detailed proof will be provided in another paper. See Pauli (29) for insightful discussion on the Inverse Square Law and Gauss' Law. Gauss' Law states that for a potential Φ(r) which is harmonic, the flux $F(S)$ of the force field $\nabla\Phi(r)$ over a the boundary $S$ of a compact n-dimensional manifold $W$ is equal to $\kappa G M$ where $\kappa$ is the surface area of the unit sphere in dimension $n$ and $M$ is the total mass enclosed by $S$. For our heuristic argument, first we assume a particle of mass $M$ is at the origin and compute the Newtonian gravitational flux of $M$ through a sphere $S$ of radius $r$ surrounding $M$. Let $W$ denote the solid ball of radius $r$ whose boundary is $S$. The Newtonian flux through $S$ is

$$F_{Newton}(S) = \oint_{S=\partial W} \nabla \Phi_{Newton}(r) \bullet dS = \int_S \frac{GM}{r^2} r^2 d\Omega = 4\pi GM \qquad (S13)$$

where $\Omega$ is the solid angle of the unit sphere in 3D and $r^2 d\Omega = dS$ is the infinitesimal area element. For the Newtonian $1/r^2$ force law, the $r^2$ terms in top and bottom of Eq. (S13) cancel to give $4\pi GM$ as the flux of the gravitational field through $S$. But for the $\log(r)$ potential, this cancellation fails and we get, instead,

$$F(S) = \oint_{S=\partial W} \nabla \Phi(r) \bullet dS = \int_S \frac{\Gamma M}{r} r^2 d\Omega = \int_S \Gamma M \, r \, d\Omega = 4\pi r \Gamma M. \qquad (S14)$$

But this has some serious consequences. For Eq. (S13), the Newtonian flux is a constant for any closed surface $S$ enclosing the mass $M$. This is no longer true for Eq. (S14) because the flux $F(S) = \int_S \Gamma M \, r \, d\Omega$ now depends on r. For example, in Eq. (S14), the flux for a sphere of radius $r$ is $4\pi r \Gamma M$, but the flux for a sphere of radius $2r$ is $8\pi r \Gamma M$. Hence the flux of the $\log(r)$ potential in 3D depends on the different shapes and sizes of the enclosing surface $S$. It also depends on the location of the particle $M$. While this is annoying and inconvenient, it also suggests a possible way to estimate the distribution of matter inside a galaxy by computing and using the flux $F(S)$ in some clever way. One possible approach for new measurements is to note that by Gauss' Divergence Theorem, we have that

$$F(S) = \oint_{S=\partial W} \nabla \Phi(r) \bullet dS = \int_W \Delta \Phi(r) dW. \qquad (S15)$$

Since $\Phi(r) = \log(r)$ is not harmonic in 3D, $\Delta \log(r) \neq 0$, so the integral of the divergence $\nabla \bullet \nabla \Phi(r)$ over the volume $W$ is no longer 0 even when the volume $W$ does not contain any sources (mass in our case). Thus, even at the far edge of the galaxy as we are on Earth, we still can measure a non-zero $F(S)$ which may yield some information about the mass distribution within our Milkyway Galaxy. Or by measuring $F(S)$ in a small volume in the direction of another galaxy like M31, we may be able to gain information about its mass distribution.

## 6. References

All references in the Supporting Information Appendix refer to the reference list of the main paper.

7. Lambers (13), Assignment 3: Assignment Nbody, from http://stanford.edu/class/cme212/, 2008.

*Assignment 3* 1

# Assignment N-body: The gravitational N-body problem

## The N-body problem

### Governing equations

Newton's law of gravitation in two dimensions states that the force exerted on particle $i$ by particle $j$ is given by

$$\mathbf{f}_{ij} = -\frac{Gm_i m_j}{r_{ij}^2}\mathbf{r}_{ij},$$

where $G$ is the gravitational constant, $m_i$ and $m_j$ are the masses of the particles, $r_{ij}$ is the distance between the particles, and $\mathbf{r}_{ij}$ is the vector that gives the position of particle $i$ relative to particle $j$. If $\mathbf{e}_x$ and $\mathbf{e}_y$ are unit vectors in the $x$ and $y$ directions, respectively, then

$$\mathbf{r}_{ij} = (x_i - x_j)\mathbf{e}_x + (y_i - y_j)\mathbf{e}_y,$$

so that

$$r_{ij}^2 = (x_i - x_j)^2 + (y_i - y_j)^2.$$

Given a distribution of $N$ particles, a straight-forward calculation of the force exerted on particle $i$ by the other $N-1$ particles is given by (using C-syle indexing)

$$\mathbf{F}_i = -Gm_i \sum_{j=0, j\neq i}^{N-1} \frac{m_j}{r_{ij}^2}\mathbf{r}_{ij},.$$

Note that by using another formula for the pairwise forces, other types of particle systems governed by Newtonian mechanics can be modeled, e.g in electrodynamics and molecular dynamics.

Using the *symplectic Euler*[1] time integration method, the velocity $\mathbf{u}_i$ and position $\mathbf{x}_i$ of particle $i$ can then be updated with

$$\begin{aligned}
\mathbf{a}_i^n &= \frac{\mathbf{F}_i^n}{m_i}, \\
\mathbf{u}_i^{n+1} &= \mathbf{u}_i^n + \Delta t \mathbf{a}_i^n, \\
\mathbf{x}_i^{n+1} &= \mathbf{x}_i^n + \Delta t \mathbf{u}_i^{n+1},
\end{aligned}$$

where $\Delta t$ is the time step size and $\mathbf{a}_i$ is the acceleration of particle $i$. This straightforward solution is prohibitively expensive for large values of $N$, since the total number of operations required to compute the forces on all $N$ particles at each time step grows as $\mathcal{O}(N^2)$.

---

[1] The symplectic Euler method is a version of the explicit Euler scheme which is the most basic example of a partitioned Runge-Kutta method (PRK). For the standard Euler scheme, the formula for $\mathbf{x}_i^{n+1}$ would read $\mathbf{x}_i^{n+1} = \mathbf{x}_i^n + \Delta t \mathbf{u}_i^n$. The symplectic Euler scheme preserves global properties of the gravitational system (e.g. total enery), while the standard Euler does not.

## Problem Setting

In this assignment you will implement a code that calculates the evolution of $N$ particles in a graviational simulation. You will calculate the motion of an initial set of particles that approximates the evolution of a galaxy. The $L \times W$ dimensionless domain (Use $L = W = 1$ in your simulations) has $N$ particles with initial positions and masses given by

$$\begin{aligned} m_i &= 1, \\ x_i &= \frac{L}{2} + R_i \cos(\theta_i), \\ y_i &= \frac{W}{2} + \alpha R_i \sin(\theta_i), \end{aligned}$$

where $R_i$ is a uniformly distributed random number between 0 and $L/4$ and $\theta_i$ is a uniformly distributed random number between 0 and $2\pi$. The parameter $\alpha$ defines the ellipticity of the initial distribution, and $\alpha = 0.25$. The initial velocities of the particles is given by solid body counterclockwise rotation such that

$$\begin{aligned} u_i &= -V R'_i \sin(\theta_i), \\ v_i &= V R'_i \cos(\theta_i), \end{aligned}$$

where $V = 50$ and $R'_i = \sqrt{(x_i - L/2)^2 + (y_i - W/2)^2}$. With $G = 0.05$, $N = 2500$, and $\Delta t = 10^{-4}$, the evolution of this initial distribution is shown in Figure 1.

## Getting started

On the Assignments page of the course web site, you can find a gzipped archive named `assignment3calving.tar.gz` for use on `calving`, or `assignment3elaine.tar.gz` for use on the `elaine`. Download this archive and unpack it. The archive contains the file `bounce.c` which should help you get started with the graphics routines that will help you debug. You can compile this program by typing `make bounce`.

The `bounce` program uses X11 graphics, so you need an X display to use it. If you do not already have an X server running, you can download one from

   http://sourceforge.net/projects/xming.

Then, you need to configure your SSH client to allow X11 forwarding. In SecureCRT, this is accomplished by opening the Session Options, and choosing the "Remote/X11" option under Connection and Port Forwarding. Check the box labeled "Forward X11 packets". In PuTTY, check "Enable X11 forwarding" on the X11 option under Connection, SSH in the PuTTY Configuration window. If you are using the given code on `elaine`, you will likely need to set your `DISPLAY` environment variable to your local IP address, followed by `:0.0` before you run `bounce`. On `elaine`, which uses `tcsh` by default instead of `bash`, the command would look something like this:

   `setenv DISPLAY 171.66.17.52:0.0`

where the IP address **must** be changed to match yours.

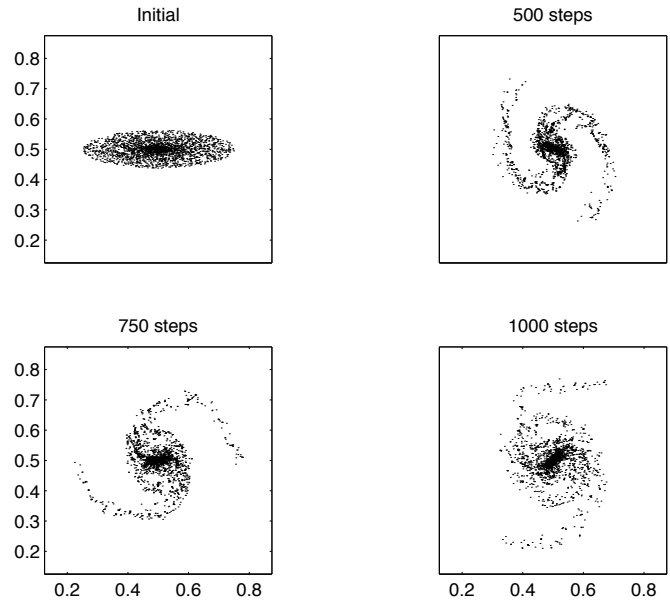

Figure 1: Evolution of a galaxy with 2500 stars after 0, 500, 750, and 1000 Euler time steps.

## Choice 1: Standard evaluation of the forces

**If you do this (easier) version of the assignment, you can pass the assignment but you will not be able to get any extra credit points.**

You should write code that implements the standard $N^2$ algorithm for computing the force. You should present a study of the effects of N on the timing of your galactic simulation over 10 time steps. You should list the results for $N = 100, 200, 400, 800, 1600, 3200$. The time should be on a per time step basis, so if the above test is performed over 10 time steps, then the time output is the average amount of time your routine took per time step over the 10 time steps, **NOT** including calls to graphics routines.

## Choice 2: Divide-and-conquer (Barnes-Hut) evaluation of the forces

**If you do this (a little more involved) version of the assignment, you can get extra credit points.**

For many problem settings, the number of operations required for computing the force in the N-body problem can be substantially reduced by taking advantage of the idea that the force exerted by a group of objects on object $i$ can be approximated as the force exerted by one object with mass given by the total mass of the group located at the center of gravity of the group of objects, as shown in Figure 2. This is a reasonable approximation as long as $h/r << 1$, which implies that the group of objects is sufficiently far away such that they behave as one object.

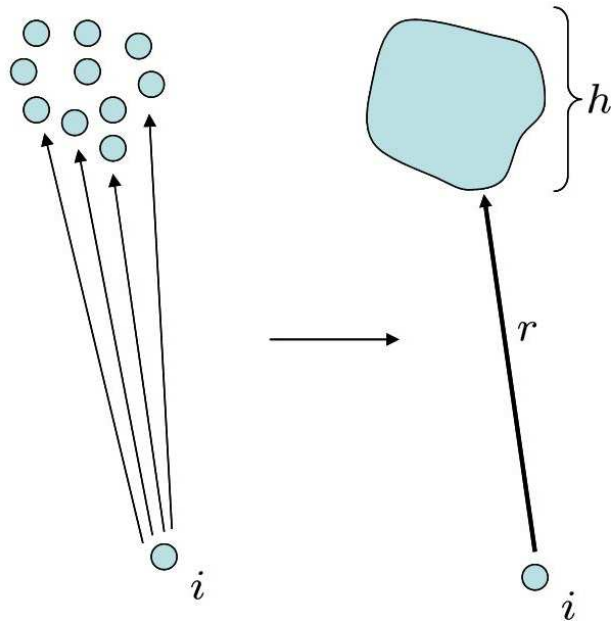

Figure 2: Approximating a group of objects as one object that is sufficienctly far away, i.e. $h/r << 1$.

This reduces the number of operations because if there are $M$ objects in the group, then $M - 1$ fewer forces need to be calculated in order to determine their net effect on object $i$. The way to handle this algorithmically is to store the particles in a quadtree and compute the forces on each object recursively. The idea behind a quadtree is to recursively subdivide the domain into four quadrangles, and subdivide those quadrangles into further sub-quadrangles, and so on, until only one object remains in each quadrangle, as shown in Figure 3.

The specific algorithm that employs the quadtree to compute the forces on a group of objects is known as the Barnes-Hut algorithm. Specifically, the Barnes-Hut algorithm involves the following high-level steps:

1. Build the quadtree for the particles or objects

2. Recursively compute the mass and center of mass of each quadrangle by adding up all of the masses contained within that quadrangle.

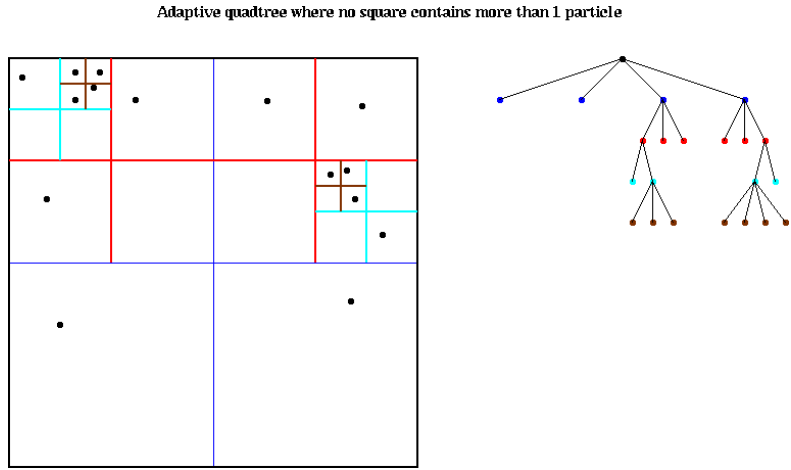

Figure 3: Adaptive grid in which each object is stored in its own quadrangle, and its associated quadtree.

3. Compute the forces on each object by recursively traversing the quadtree.

The key ingredient of the Barnes-Hut algorithm is to decide whether or not to traverse down the branches of a node and compute the forces resulting from the particles contained within its branches, or just to stop at the current node and assume all of its particles can be lumped into an equivalent mass at its center of mass. This is determined by the threshold $\theta$, which is usually less than 1, and is defined as

$$\theta = \frac{\text{width of current box containing particles}}{\text{distance to particle}}.$$

You should write code that implements the Barnes-Hut algorithm and present a study of the effects of $\theta$ and N on the timing of your galactic simulation over 10 time steps. You should list the results of $N = 100, 200, 400, 800, 1600, 3200$, with $\theta = 0.1, .2, .4, .8, 1.6$. The time should be on a per time step basis, so if the above test is performed over 10 time steps, then the time output is the average amount of time your routine took per time step over the 10 time steps, **NOT** including calls to graphics routines.

## Report

Regardless of which algorithm you implement, you must to write a report describing your project. The report shall contain the following sections:

- The Problem

- The Solution (Describe the data structures, the structure of your code, and how the algorithms are implemented. Are there other options, and why did you not use them?)

- Experiments, optimization and discussion. (Present experiments where you investigate the performance of the algorithm and your code)

- A listing of the code (as an appendix).

The report should either be a PDF file or a Word document. The discussion to be held on February 21 will cover LaTeX and writing reports.

## Deadline

To submit your code and report, run the `submit` script on `calving` as follows,

`/home/lambers/cme212/submit 3` *filename*

from the directory in which your project files are stored, where *filename* is the name of your report file. In this directory, there should be a `Makefile` which builds your program on `calving` when typing `make`. Also, create a `README` file where you explain how to use your program, and place it in the same directory as your source files and `Makefile` before running the `submit` script. If you are working on `elaine`, use this command to submit:

`/usr/class/cme212/submit 3` *filename*

**The due date is 11:59pm on Tuesday, February 26, 2008**.

8. Lambers (14), Questions Page (On error in Assignment 3), Feb 14, Feb 18, Feb 22, 2008, http://stanford.edu/class/cme212/.

This documents the error in Assignment 3 where Newton's $1/r^2$ force was mistakenly set to $1/r$ as noted by the students and the instructor's exchange. See notes on Feb 14, 18, 22. These are screen dumps from the course website:

| | | |
|---|---|---|
| Feb 18 | the assignment reads "the time output is the average amount of time your routine took per time step over the 10 time step". Why do we divide twice by the number of time step ? Don't we just have to take the average ? | There is only one division; "average" and "over" refer to that same division. Add up the total amount of time taken, and divide it by the number of time steps. |
| Feb 18 | considering the expression we have for yi in the Problem Setting, don't we have for the velocity : yi = alpha * V * Ri * cos(Oi) ? | Yes there should be an alpha there. |
| Feb 17 | 1. Which are the extra credit options? The assignment requires us to implement the standard method, as base case and the Barnes-Hut method for the extra credit option. In class, I remember your mentioning other methods e.g., Cut-off etc., which are also in the hand out slides for project 3.<br>2. In the updating the positions for the particles and computing the forces, do we use synchronous vs. asynchronous update i.e., jacobi- vs. gauss-seidel-type update for the particles. I am not sure, but my guess will be the gauss-seidel type update (where each particle will use the recently computed positions of other particles from the current time step), may yield a better and more realistic simulation. | 1. Those other methods mentioned in the slides are not part of the project. There's only the naive implementation and Barnes-Hut. Cut-off, in particular, is not an accurate method anyway.<br>2. There are advantages and disadvantages to either approach, but I would suggest using the Jacobi-type updating since that's what is prescribed in the handout. |
| Feb 15 | there must be something really simple i'm missing here:<br><br>i've taken the original bounce.c code and just changed a few lines to start using cosine and sine functions,<br><br>namely "cos(theta[i])" and the like... but when i type make, i get undefined reference to 'cos'. i checked that bounce.c includes <math.h> or "math.h", neither works<br><br>and in the makefile, there is a line that says bounce.o: /usr/include/math.h<br><br>so it seems like i have the references right, by why isn't make finding it? | You need to link with the math library, using the -lm option. |
| Feb 14 | I think there is an error in the handout for Assignment 3. On the first formula, rij should be to the power of 3 and not 2, since you multiply the whole expression by vector rij. | Yes, it should be 3. I'll let Henrik know about this. |

Project 2

## II. Supporting Movies S1-S4 Legends

Animations at: **http://martinlo.com/Home/RNBP.html**

Movie S1. Title: Cartwheel Galaxy & 1/r Gravity Law

    File: LoMovieS1.mp4               (9.7 MB)
Summary: This animation shows the simulations of the Cartwheel Galaxy. The particles are colored by their initial distance from the center of the galaxy as indicated in Frame 0 of the animation. This is in order to track the particles and see what kind of mixing takes place. Rings are generated by density waves from the center. Multiple rings are connected by spokes producing the cartwheel effect, hence the name. The animation divides the galaxy into bands of particles and to show how the bands evolve. The bands of particles are not mixing from one band to the other.

Movie S2. Title: Spiral Galaxy & 1/r Gravity Law

    File: LoMovieS2.mp4               (8.7 MB)
Summary: This animation shows the simulations of the the Spiral Galaxy. The particles are colored by their initial distance from the center of the galaxy as indicated in Frame 0 of the animation. This is in order to track the particles and see what kind of mixing takes place. The Spiral arms are formed by density waves which are incomplete rings. This produces bands and smaller rings which survive. This animation shows spiral galaxies as incomplete cartwheel ring galaxies.

Movie S3. Title: Uniform Galaxy & 1/r Gravity Law

    File: LoMovieS3.mp4               (7.9 MB)
Summary: This animation shows the evolution of a set of particles uniformly distributed in Cartesian coordinates in a circular region. Interestingly, the particles first clump and from filaments and voids which resemble the large scale structure of the universe. It then evolves into a galaxy with a broken center with multiple arms. It is trying to form a center but never succeeds until it finally evolves into an elliptical galaxy.

Movie S4. Title: Galaxy Collision & 1/r Gravity Law

    File: LoMovieS4.mp4               (7.2 MB)
Summary: This animation shows the collision of two galaxies using the galaxies from Case A and Case B with slightly less number of particles. After repeated collision encounters, the two galaxies merge to form an elliptical galaxy with a bar in the center.

**Acknowledgments**